\documentclass[10pt, a4paper]{article}
\usepackage[utf8]{inputenc}
\usepackage[margin=1.5cm]{geometry}
\usepackage{authblk} % Essencial para formatar múltiplos autores/afiliações
\usepackage[colorlinks=true, allcolors=blue]{hyperref} % Links clicáveis no PDF

\providecommand{\doi}[1]{\href{https://doi.org/#1}{#1}}

\usepackage{amsmath}
\usepackage{amssymb}
\usepackage{bm}
\usepackage{wasysym} % Símbolos astrológicos
\usepackage{mathrsfs} % Fonte matemática estilizada \mathscr{}
\usepackage{cancel}  % Para o comando \cancel

\usepackage{graphicx}
\usepackage{subcaption} % Subfiguras
\usepackage{multirow}
\usepackage{booktabs} % Tabelas profissionais (\toprule, \midrule)
\usepackage{float} % Opção [H] para tabelas e figuras
\usepackage{placeins} % Comando \FloatBarrier
\usepackage[round, authoryear]{natbib}

\usepackage[table]{xcolor}
\definecolor{lightgray}{gray}{0.9}
\definecolor{newcolor}{rgb}{.8,.349,.1}

\usepackage{tikz}
\usepackage{pgfplots}
\pgfplotsset{compat=newest} % Garante a versão mais recente do pgfplots

\usepackage{url}
\usepackage{lineno} % Mantido caso queira numeração de linhas nas revisões

\usepackage{newtxtext, newtxmath}
\usepackage{booktabs}

\title{A Semi-Analytical Theory for Improved Orbit Design of the GARATÉA-L}

\author[1]{Luiz Arthur Gagg Filho\thanks{Corresponding author: e-mail: luiz.gagg@gp.ita.br}}
\author[2]{Sandro da Silva Fernandes}

\affil[1]{\small  Flight Mechanics Department, Instituto Tecnológico de Aeronáutica, Praça Marechal Eduardo Gomes 50, Vila das Acácias, 12228-900, São José dos Campos-SP, Brazil.}
\affil[2]{\small  Mathematics Department, Instituto Tecnológico de Aeronáutica, Praça Marechal Eduardo Gomes 50, Vila das Acácias, 12228-900, São José dos Campos-SP, Brazil.}

\date{} % Deixe vazio para não imprimir a data de hoje, ou remova para imprimir

\begin{document}
	
	% Na classe article, \maketitle precisa ser chamado dentro do document
	\maketitle
	
	% Removemos o \begin{frontmatter} da Elsevier. O abstract vem logo após o \maketitle.
\begin{abstract}
	This work proposes a semi-analytical theory for improving orbit design of the GARATÉA-L Brazilian lunar mission. The dynamical model incorporates the lunar gravitational potential up to degree 12 and order 3 harmonics and a third-body perturbation model accounting for the eccentricity and inclination of the Earth's orbit with respect to the Moon. Through Hori's method, short- and medium-period terms are eliminated from the Hamiltonian to derive a first-order theory in mean orbital elements. It is demonstrated that nominal frozen orbits designed under simplified circular and equatorial assumptions for the motion of the third body are not maintained in this more realistic dynamical environment, leading to significant eccentricity growth and potential collisions of the probe with the lunar surface.  The inclusion of the third body's inclination and eccentricity breaks the system's axial symmetry, introducing the longitude of the ascending node into the double-averaged variational equations and rendering stationary frozen orbits practically unattainable. Consequently, the problem is addressed through a parametric study around the simplified model's frozen conditions to identify ``low-drift" states where eccentricity variations are highly decreased. A new nominal orbit is proposed for GARATÉA-L. Numerical validation against high-fidelity propagation and JPL ephemeris confirms that this new configuration maintains the ``low-drift" behaviour throughout the mission, ensuring orbital safety and avoiding the collision predicted by simpler models.
\end{abstract}

% Delete tudo que for \begin{keyword} e \end{keyword} e substitua por:
\vspace{0.5cm}

%\begin{keyword}
%	
%	GARATÉA Brazilian probe \sep zonal harmonics effects \sep third-body effects \sep Hori's method \sep lunar mission
%	%% keywords here, in the form: keyword \sep keyword
%	
%	%% MSC codes here, in the form: \MSC code \sep code
%	%% or \MSC[2008] code \sep code (2000 is the default)
%	
%\end{keyword}  

%\linenumbers


\section{Introduction}

The dynamical environment in the vicinity of the Moon presents a unique set of challenges for astrodynamics, distinct from the well-known ``Earth-type'' problems dominated by the second zonal harmonic $J_2$ \citep{depritEarthtype, brouwer1959}. In the lunar case, the gravitational potential is characterized by a non-central and highly asymmetric field where no single spherical harmonic coefficient holds absolute dominance. As discussed by \cite{desaedeleer2004} and \cite{konopliv2001}, the Moon's slow synchronous rotation and the presence of mass concentrations require the simultaneous treatment of higher-order harmonics to accurately model the motion of artificial satellites. Furthermore, for high-altitude orbiters, the third-body perturbation exerted by the Earth becomes a governing force, often surpassing the effects of the lunar non-sphericity \citep{ely2005, folta2006}.

The fundamental dynamics of a satellite subject to a distant third-body perturbation were established in the works of \cite{lidov-kozai} and \cite{kozai1962}. Lidov demonstrated that for satellites with high apogees, the gravitational influence of external bodies (such as the Moon or Sun acting on an Earth satellite) can become comparable to the central body's oblateness, driving secular evolution that may lead to a rapid increase in eccentricity and subsequent collision. Simultaneously, Kozai provided the analytical framework for asteroids with high inclination and eccentricity, revealing the coupling between eccentricity and inclination —now recognized as the Lidov-Kozai resonance. For lunar polar orbiters, this mechanism is particularly critical, as the interaction between the Earth's pull and the lunar potential can destabilize the orbit if not properly mitigated.

With the recent resurgence of interest in lunar exploration, driven by the Artemis program and the proliferation of low-cost platforms such as CubeSats, the demand for robust analytical and semi-analytical propagation models has grown significantly. Unlike traditional large spacecraft, these CubeSats often have limited propellant budgets, making the identification of naturally stable or ``frozen'' orbits a critical design requirement. Building on the classical zonal problem, \cite{elipe2003} applied averaging techniques to identify families of frozen orbits around the Moon. More recently, \cite{singh2020} conducted an extensive feasibility study on quasi-frozen orbits at altitudes below 100 km, emphasizing that for these missions, high-fidelity gravity models are a necessity to predict mission lifetimes.

Two general perturbation methods have historically been applied to derive formal solutions for this problem: the Von Zeipel method, utilized in the work of \cite{giacaglia1970}; and Lie Transform methods, such as those proposed by \cite{hori1966} and \cite{deprit1969}. With the advancement of symbolic manipulation software, recent studies have extended these theories to higher degrees of complexity. \cite{mastroianni2025} proposed a fully analytical propagator in closed form, eliminating the need for series expansions in eccentricity. Their approach is particularly relevant for orbits spanning the 300 to 3000 km range, achieving an accuracy comparable to high-order semi-analytical propagators while maintaining the speed of a fully analytical solution.

Despite these advances, a recurring limitation in many semi-analytical models is the simplification of the third-body geometry. Several theories still rely on a circular-equatorial approximation for the perturbing body's orbit. While this assumption facilitates the elimination of the explicit time dependence in the Hamiltonian, it fails to capture critical secular effects, particularly for high-altitude missions. For instance, \cite{nie2019} utilized the von Zeipel method to construct a semi-analytical model accounting for the inclination and eccentricity of the third body, demonstrating that this geometrical sophistication has a vital influence on the long-term evolution of the satellite's orbit. As such, the present work focuses on incorporating the inclination and eccentricity of the third body into the semi-analytical framework.

Leveraging the advantages of applying perturbation theory for mission design—specifically the ability to extract global dynamical features and reduce computation time—this paper develops an improved semi-analytical theory for the GARATÉA-L mission. GARATÉA-L is a Brazilian lunar CubeSat mission designed to investigate the biological effects of deep-space radiation on extremophile microorganisms \citep{garatea_revistaFapesp}. The mission targets a highly inclined, elliptical orbit, making it particularly sensitive to the coupling between high-order zonals and the Earth's orbital geometry. 

In a preliminary study presented at the COBEM congress, \cite{gaggCOBEM2025} applied Hori's method using a simplified equatorial third-body model to design the GARATÉA-L orbit. However, as noted in that work, comparisons with numerical ephemeris propagation revealed that the identified frozen conditions were not maintained when the Earth's orbital eccentricity and inclination were included. The present work extends and improves the theory proposed in \cite{gaggCOBEM2025} and \cite{gagg2025} by incorporating a $12 \times 3$ lunar gravity model and a comprehensive third-body potential that accounts for both the inclination and eccentricity of the Earth's orbit. Hori's method is applied successively to eliminate short- and medium-period terms, deriving a system of Lagrange equations for the mean elements. This enhanced model allows for the identification of ``low-drift'' orbits—solutions where the eccentricity and argument of pericenter varies with minimal amplitude. The results are validated against high-fidelity numerical propagation with a $50 \times 50$ harmonic model and JPL ephemerides, demonstrating that incorporating the Earth's true orbital parameters is essential for the accurate prediction of the orbital lifetime of the GARATÉA-L probe.

The remainder of this article is organized as follows. Section \ref{sec:formulation} establishes the general mathematical framework, detailing the reference systems, the expansion of the lunar gravitational potential, and the development of the eccentric-inclined third-body perturbation. Since the  mathematical  structure follows the work in \cite{gagg2025}, particular emphasis is placed on the third-body generalization introduced here. This section also describes the sequential application of Hori’s method to eliminate short- and medium-period terms from the Hamiltonian to derive the mean equations of motion. It concludes with a truncation theory analysis based on the established dynamical hierarchy. 
Section 3 presents the results and discussion, beginning with the primary motivation for this research: the inability of previous simplified models to accurately predict orbital evolution, particularly regarding the unforeseen collisions observed in ephemeris-based models. After demonstrating that these discrepancies stem from the circular-equatorial simplification of the third body, the section details the parametric study performed to identify ``low-drift" orbit conditions for the GARATÉA-L mission using the semi-analytic truncated theory. This analysis specifically accounts for the broken axial symmetry of the system to ensure mission safety. The findings are then validated through a comprehensive comparison between the semi-analytical theory, high-fidelity numerical models, and JPL ephemerides. Finally, Section 4 summarizes the main conclusions and provides perspectives for future developments.

\section{Mathematical General Model}
\label{sec:formulation}

This section establishes the analytical framework governing the orbital motion of a lunar probe. The approach relies on the Hamiltonian formalism combined with Hori's perturbation method \citep{hori1966}, a canonical transformation theory based on Lie series. The primary objective is to derive a mean Hamiltonian free of short-period terms (related to the satellite's mean anomaly) and medium-period terms (related to the Moon's rotation). From this averaged Hamiltonian, Lagrange planetary equations is derived and solved numerically to obtain the long-term evolution of orbital elements. Furthermore, short and medium-period variations are analytically recovered via Poisson brackets involving the generating functions.

\subsection{Reference Systems and Dynamical Model}

\begin{figure}
	\centering
	\includegraphics[width=0.5\linewidth]{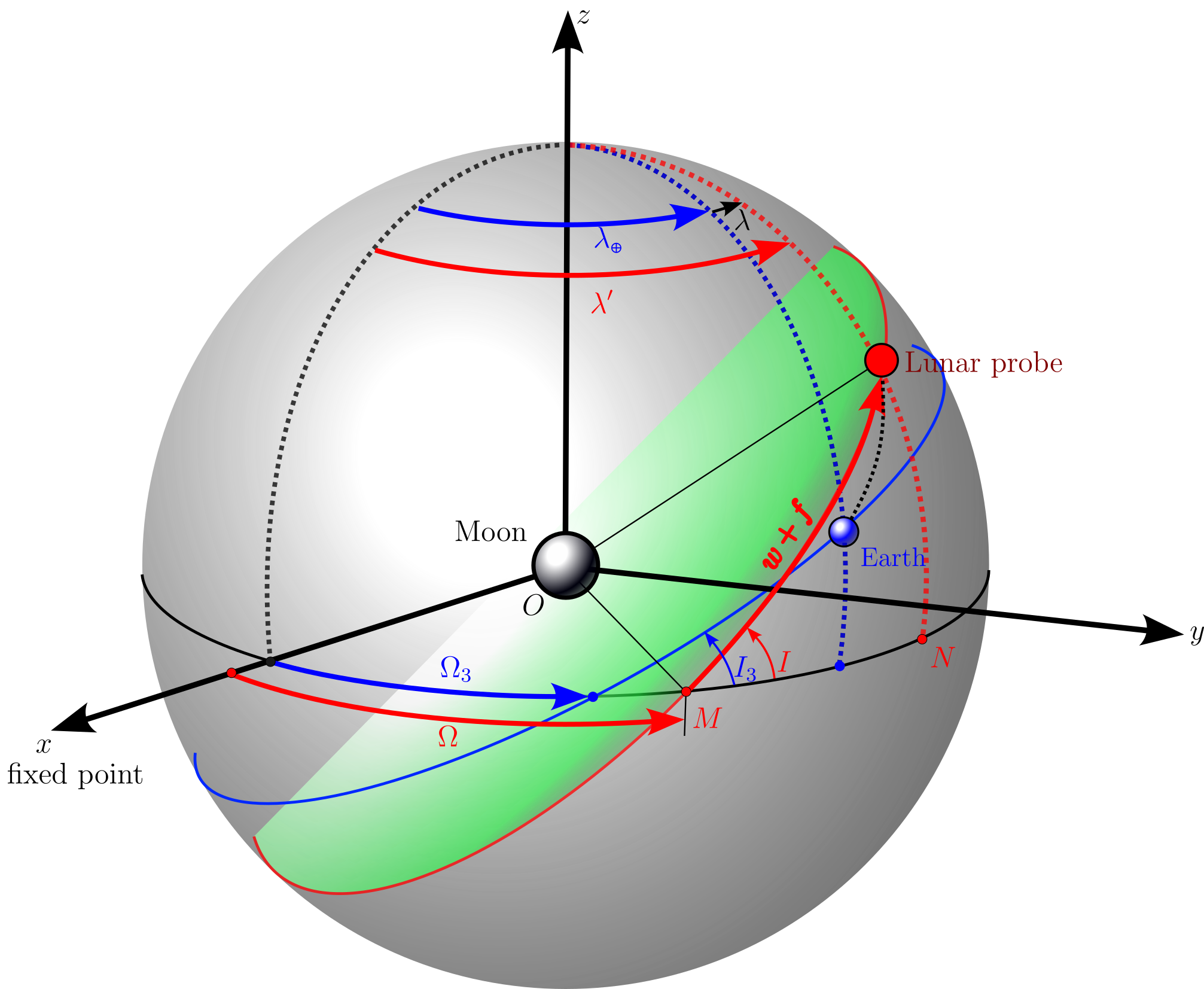}
	\caption{Geometry and reference frame.}
	\label{Fig.1}		
\end{figure}

The dynamical model considers the motion of a lunar orbiter subject to the non-spherical gravitational field of the Moon and the gravitational attraction of the Earth as a third-body perturbation. 
To describe the system, two reference frames are defined:

\begin{enumerate}
	
	\item \textbf{The Inertial Frame ($Oxyz$):} The origin $O$ is located at the Moon's center mass. The fundamental plane ($xy$) coincides with the lunar mean equatorial plane. The $Ox$ axis points towards a fixed point (vernal equinox), and the $Oz$ axis is aligned with the Moon's rotation axis/ angular momentum, as illustrated in Fig. \ref{Fig.1}.
	
	\item \textbf{The Selenocentric Rotating Frame ($Ox'y'z'$):} Also referred to as the ``mean Earth/polar axis'' system, this frame rotates with the Moon. The $Ox'$ axis points toward the mean direction of the Earth, and the frame rotates around the $Oz$ axis with a constant angular velocity $\omega_{\leftmoon}$, corresponding to the lunar mean synchronous rotation rate.

\end{enumerate}

The relationship between the frames is governed by the rotation angle $\theta = \omega_{\leftmoon} t + \theta_0$, which connects the position vector $\mathbf{r}$ in both systems via a standard transformation matrix. 	

While the fundamental canonical structure follows the framework previously established in \cite{gagg2025} and \cite{gaggCOBEM2025}, the present formulation introduces a significant generalization regarding the third-body perturbation. Unlike prior simplified models that restricted the disturbing body to a circular-equatorial path in the $xy$ plane, the current derivation accounts for the geometry of the apparent Earth orbit. Here, the Earth follows a Keplerian orbit defined by its semi-major axis $a_\oplus$, eccentricity $e_\oplus$, inclination $I_\oplus$, longitude of the ascending node $\Omega_\oplus$, and argument of pericenter $\omega_\oplus$, all measured with respect to the lunar equatorial plane. Consequently, the position vector $\mathbf{r}_\oplus$ is a time-dependent function governed not only by the mean anomaly but also by the spatial orientation of the orbital plane. As will be demonstrated, this generalization breaks the axial symmetry of the third-body potential, introducing a dependency of the mean elements on the longitude of the ascending node $\Omega$. This shift redefines the conditions for frozen orbit —a decisive factor for the long-term mission analysis of high-altitude lunar orbiters.

%\begin{figure}
%\begin{subfigure}{0.5\linewidth}
%	\centering
%	\includegraphics[width=1.0\linewidth]{Figuras/SR.png}		
%\end{subfigure}
%\hfill
%\begin{subfigure}{1.0\linewidth}
%	\centering
%	\includegraphics[width=0.5\linewidth]{Figuras/SR2.png}	
%	\caption{Spherical triangle PMN.}	
%\end{subfigure}	
%\caption{Reference systems.}
%\label{Fig.1}
%\end{figure}

\vspace {1cm}

\subsection{Lunar gravitational potential}

The gravitational potential of the Moon is represented by the standard expansion in spherical harmonics \citep{kaula1960}:

\begin{equation}
	V=-\frac{G m_{\leftmoon}}{r}\left[1-\sum_{n=2}^\infty J_{n}\left(\frac{a_{\leftmoon}}{r}\right)^n P_{n}\left(\sin \phi\right)+\sum_{n=2}^\infty \sum_{m=1}^n  \left(\frac{a_{\leftmoon}}{r}\right)^n P_{nm}\left(\sin \phi\right)\left(C_{nm} \cos \left(m\lambda\right)+S_{nm} \sin \left(m\lambda\right)\right)\right]
	\label{eq.1}
\end{equation} 

\noindent where $m_{\leftmoon}$ is the mass of the Moon, $G$ is the universal gravitational constant, $a_{\leftmoon}$ denotes the mean equatorial radius of the Moon ($1738$ km),  $P_n$ is Legendre polynomial of degree $n$, $P_{nm}$ is associated Legendre function of degree $n$ and order $m$, $\left(r,\phi,\lambda\right)$ are spherical coordinates with $r$ standing for selenocentric radius, $\phi$ for latitude, $\lambda$ for longitude, $J_n$ is the coefficient for the zonal harmonic, $C_{nm}$ and $S_{nm}$ are the coefficients for tesseral $(n=m)$ or sectorial $(n \ne m)$ harmonics. The most relevant coefficients adopted in this work for the Moon's gravitational model are presented in Tab. 1 \citep{konopliv2001}.

%The gravity acceleration is given by $\mathbf g=-\nabla \Phi$.

\begin{table}[h]	
	\centering
	\caption{ Coefficients for zonal, tesseral and sectorial harmonics of the lunar gravitational potential}
	%	\vspace{0.5 cm}
	\begin{tabular}{c c c c}
	\toprule
	\textbf{Coefficient} & \textbf{Value} & \textbf{Coefficient} & \textbf{Value} \\
	\midrule
	$J_2$    & $+2.032365 \times 10^{-4}$ & $C_{22}$ & $+2.235491 \times 10^{-5}$ \\
	$J_3$    & $+8.585355 \times 10^{-6}$ & $S_{22}$ & $+1.535185 \times 10^{-8}$  \\
	$J_4$    & $-9.860033 \times 10^{-6}$ & $C_{31}$ & $+2.849956 \times 10^{-5}$ \\
	$J_5$    & $+8.393511 \times 10^{-7}$ & $C_{32}$ & $+4.847437 \times 10^{-6}$ \\
	$J_6$    & $-1.331149 \times 10^{-5}$ & $C_{33}$ & $+1.712582 \times 10^{-6}$ \\
	$J_7$    & $-2.205653 \times 10^{-5}$ & $S_{31}$ & $+5.953103 \times 10^{-6}$ \\
	$J_8$    & $-9.434727 \times 10^{-6}$ & $S_{32}$ & $+1.689177 \times 10^{-6}$ \\
	$J_9$    & $+1.511956 \times 10^{-5}$ & $S_{33}$ & $-2.564964 \times 10^{-7}$ \\
	$J_{10}$ & $+3.863970 \times 10^{-6}$ &          &                            \\
	$J_{11}$ & $+5.130603 \times 10^{-6}$ &          &                            \\
	$J_{12}$ & $+9.103736 \times 10^{-6}$ &          &                            \\
	\bottomrule
\end{tabular}
	\label{Tab1}
\end{table}

In order to build a semi-analytic theory, a set of classical orbital elements $\left( a,e,I,\Omega, \omega, f \right)$ is introduced, where $a$ stands for semimajor axis, $e$ for eccentricity, $I$ for inclination of the orbit plane, $\Omega$ for longitude of ascending node, $\omega$ for argument of pericenter and $f$ for true anomaly. For completeness, a brief description of this development is presented in the next paragraphs. The alternative procedure to compute Kaula's inclination function for the terms of the potential associated to the tesseral and sectorial harmonics was already described in \cite{gagg2025}.

The first term in Eq. \ref{eq.1},

\begin{equation*}
	V_0=-\frac{G m_{\leftmoon}}{r},
\end{equation*}

\noindent governs the unperturbed Keplerian motion. Focusing first on the zonal harmonics ($m=0$), the sum of the corresponding potential terms is given by:
\begin{equation*}
	V_z= \sum_{n=2}^\infty V_n
\end{equation*} 
where
\begin{equation*}
	V_n=\frac{G m_{\leftmoon}}{r} J_{n}\left(\frac{a_{\leftmoon}}{r}\right)^n P_{n}\left(\sin \phi\right)
\end{equation*} 
\noindent Utilizing the explicit series representation for the Legendre polynomial $P_n(x)$ \citep{arfken1972}:

\begin{equation}	
	P_n(x)=\frac{1}{2^n} \sum_{k=0}^{\left[\frac{n}{2}\right]}\frac{\left(-1\right)^k \left(2n-2k\right)!}{k!\left(n-k\right)!\left(n-2k\right)!}x^{\left(n-2k\right)}
	\label{eq:legendre_poly}
\end{equation}

\noindent where $\left[\frac{n}{2}\right]$ denotes the integer part of $\frac{n}{2}$.  Applying the law of sines to the spherical triangle formed by the satellite's position, the ascending node, and the projection on the orbital plane (Fig. \ref{Fig.1}),the relationship $\sin \phi = \sin I \sin(\omega+f)$ is obtained. Substituting this result into Eq. (\ref{eq:legendre_poly}) allows the zonal potential to be expressed in terms of the orbital elements:

\begin{equation}
	V_n = \frac{\mu}{a}  J_n \left( \frac{a_{\leftmoon}}{a} \right)^n \sum_{p=0}^{\lfloor n/2 \rfloor} F_{n,n-2p}(I) \left(\frac{a}{r}\right)^{n+1} 
	\begin{cases}
		\cos((n-2p)(\omega+f)) & \text{if } n \text{ is even} \\
		\sin((n-2p)(\omega+f)) & \text{if } n \text{ is odd}
	\end{cases}
	\label{eq:zonal_potential}
\end{equation}

\noindent where, $\mu$ denotes Moon's gravitational parameter, $\mu=Gm_{\leftmoon}$. The inclination functions $F_{n,n-2p}(I)$ are Kaula's inclination functions \citep{kaula1960, osorio}. In this work, a recursive formulation derived in previous work \citep{gagg2025} is adopted for these functions to facilitate symbolic implementation:

\begin{equation}
	F_{n,n-2p}(I)=\sum_{j=0}^{p} \alpha_{n,j}\left(-1\right)^{p-j}\left(\begin{array} {cc} n-2j  \\ p-j \end{array}\right)(\sin I)^{n-2j}
	\label{eq.5}
\end{equation}  for $n\ne 2p$, and, 

\begin{equation}
	F(I)_{n,0} =  \sum_{j=0}^{p}  {n-2j \choose p-j}\left(\sin I\right)^{n-2j}\beta_{n,j} \quad  
	\label{eq74a}
\end{equation} for  $n = 2p$, with the coefficients $\alpha_{n,j}$ and $\beta_{n,j}$ defined as:

\begin{equation}
	\alpha_{n,j}=\left(-1\right)^{[\frac{n}{2}]} \frac{(2n-2j)!}{2^{2n-1-2j}j!(n-j)!(n-2j)!}
	\label{eq.6}
\end{equation}

\begin{equation}
	\beta_{n,j} = 
	(-1)^j \frac{\left(2n-2j \right)!}{j! (n-j)! (n-2j)!}  %\left(\sin I \right)^{2(n-k)} 
	\Bigg[
	\frac{1}{2^{ 2n-2j}} %{ 2(n-k) \choose n-k} 
	%\left\langle
	\Bigg]		
	\label{eq6a}
\end{equation}

\noindent The inclination functions $F_{n,n-2p}(I)$ were presented in Appendix of the previous paper \citep{gagg2025}.

\subsubsection{Tesseral and Sectorial Harmonics}

The contribution of tesseral ($n \neq m$) and sectorial ($n=m$) harmonics to the potential is expressed as:

\begin{equation}
	V_{t}=\sum_{n=2}^\infty \sum_{m=1}^n V_{nm}
\end{equation} 
where 
\begin{equation}
	V_{nm}=-\frac{G m_{\leftmoon}}{r} \left(\frac{a_{\leftmoon}}{r}\right)^n P_{nm}\left(\sin \phi\right)\left(C_{nm} \cos \left(m\lambda\right)+S_{nm} \sin \left(m\lambda\right)\right)
	\label{eq.7}
\end{equation} 

\noindent The associated Legendre functions $P_{nm}(x)$ relate to the standard polynomials via derivatives \citep{arfken1972}:

\begin{equation}
	P_{nm}(x)=\left(1-x^2\right)^\frac{m}{2} \frac{d^m}{dx^m}	P_n(x)
	\label{eq.8}
\end{equation}

\noindent From Eq. \ref{eq:legendre_poly}, 

\begin{equation}
	P_{nm}(x)=\left(1-x^2\right)^\frac{m}{2} \frac{1}{2^n} \sum_{k=0}^{\left[\frac{n-m}{2}\right]}\frac{\left(-1\right)^k \left(2n-2k\right)!}{k!\left(n-k\right)!\left(n-2k-m\right)!}x^{\left(n-2k-m\right)}
	\label{eq.9}
\end{equation}

By applying the spherical law of cosines to the geometry in Fig. \ref{Fig.1}, the following expansions is derived \citep{gagg2025} for the tesseral potential in orbital elements:

\begin{equation}
	V_{nm} =  -\frac{\mu}{a}\left(\frac{a_{\leftmoon}}{a} \right)^n \left(\frac{a}{r}\right)^{n+1} 	\sum_{p = 0}^{n} 
	F_{n,m,n-2p}(I)\left[ C_{nm} \cos\left((n-2p)(\omega+f)+m\left(\Omega - \omega_{\leftmoon}t\right)\right) + S_{nm} \sin\left((n-2p)(\omega+f)+m\left(\Omega - \omega_{\leftmoon}t\right)\right) 
	\right]
	\label{eq.14}
\end{equation}

\noindent if $n-m$ is even, and,

\begin{equation}
	V_{nm} =  -\frac{\mu}{a}\left(\frac{a_{\leftmoon}}{a} \right)^n \left(\frac{a}{r}\right)^{n+1} 	\sum_{p = 0}^{n} 
	F_{n,m,n-2p}(I)\left[ C_{nm} \sin\left((n-2p)(\omega+f)+m\left(\Omega - \omega_{\leftmoon}t\right)\right) - S_{nm} \cos\left((n-2p)(\omega+f)+m\left(\Omega - \omega_{\leftmoon}t\right)\right) 
	\right]
	\label{eq.15}	
\end{equation}

\noindent if $n-m$ is odd.    The symbol $\omega_{\leftmoon}$ stands for the Moon's rotation rate, and the functions $F_{n,m,n-2p}(I)$ are Kaula's inclination functions expressed as 

\begin{equation}
	F_{n,m,n-2p}(I)=\sum_{k=k_1}^{k_2}H_{n,m,n-m-2l}(I) \mathscr{C}_{m,m-2k}(I)
	\label{eq.16}
\end{equation}

\noindent where $k_1=\max\{0,m-n+p\}$, $k_2=\min\{m,p\}$, and $l = p-k$.   The functions $\mathscr{C}_{m,m-2k}$ are given by

\begin{equation}
	\mathscr{C}_{m,m-2k} (I) = {m \choose k} \left(\frac{1}{2}\right)^m \left(1+\cos I\right)^{m-k} \left(1-\cos I\right)^k
\end{equation}

\noindent and, the auxiliary inclination functions $H_{n,m,n-m-2l}(I)$ are defined as \citep{gagg2025}

\begin{equation}
	H_{n,m,n-m-2l}(I) = 
	\sum_{k=0}^{n-m-l}\alpha_{n,m,k}^\star \gamma_{s,l-k} \left(\sin I\right)^{n-m-2k} 
	\label{eq.17}
	%	\label{eq_tesse_199}
\end{equation}

\noindent if $n-m$ is even, and,

\begin{equation}
	H_{n,m,n-m-2l}(I) = 
	\sum_{k=0}^{n-m-l}\alpha_{n,m,k}^\star \tilde{\gamma}_{s,l-k} \left(\sin I\right)^{n-m-2k} 
	\label{eq.17b}
	%	\label{eq_tesse_221}
\end{equation}

\noindent if $n-m$ is odd

\begin{equation}
	\alpha_{n,m,k}^\star = 
	(-1)^k \frac{(2n-2k)! }{2^n k! (n-m-2k)! (n-k)!
		\label{eq_tesse_219}   %Observar equações 454 e 495 (Repare que são os mesmos) 
	}
\end{equation}

\noindent where 
\begin{equation}
	\gamma_{s,t} =  \frac{(-1)^{s+1}}{2^{2s+2}}(-1)^tk{2s+2 \choose t}
	\label{eq_tesse_188}
\end{equation}

\noindent and, 
\begin{equation}
	\tilde{\gamma}_{s,t} =  \frac{(-1)^{s}}{2^{2s+1}}(-1)^t{2s+1 \choose t}.
	\label{eq_tesse_213}
\end{equation}

\noindent where $s = \frac{n-m-2k-2}{2}$ and $s = \frac{n-m-2k-1}{2}$, respectively, for $n-m$ even and $n-m$ odd; and, $t = l-k$.   The inclination functions $F_{n,m,n-2p}(I)$ were also presented in Appendix of the previous paper \citep{gagg2025}.

This formulation, consistent with \cite{vallado2007} and \cite{segerman}, is optimized for symbolic manipulation. It is worth noting that for $m=0$, Eqs. \ref{eq.14} and \ref{eq.15}, reduce exactly to the zonal formulation derived in Eq. (\ref{eq:zonal_potential}), with the identification $J_n = -C_{n,0}$.

%\color{blue}

\subsection{Third body perturbation}

The disturbing potential arising from the gravitational attraction of the Earth is expressed in terms of Legendre polynomials, following the classical formulation found in \cite{cook1962} and \cite{vallado2007}:

\begin{equation}
	V_{\oplus} = -\frac{\mu_\oplus}{r_\oplus} \sum_{k=2}^{\infty} \left( \frac{r}{r_\oplus} \right)^k P_k(\cos S)
\end{equation}

\noindent where $\mu_\oplus$ is the Earth's gravitational parameter, $r_\oplus$ is the Earth-Moon distance, and $S$ denotes the angle between the selenocentric position vectors of the satellite ($\mathbf{r}$) and the Earth ($\mathbf{r}_\oplus$). In this work, the expansion is truncated at term  $k=2$, yielding: 

\begin{equation}
	V_{\oplus,{P_2}} = -\frac{\mu_\oplus}{r_\oplus	} \left( \frac{r}{r_\oplus} \right)^2 \left( \frac{3}{2} \cos^2 S - \frac{1}{2} \right)
	\label{eq:V3b_quad}
\end{equation}

\noindent A detailed justification for this truncation, based on an order-of-magnitude analysis and the dynamical hierarchy of the system, is provided in Section \ref{section_truncation}.

A distinct feature of the present formulation, which represents a improvement over previous models \citep{gagg2025, gaggCOBEM2025}, is a more complex treatment of the Earth's orbital geometry. Unlike simplified models that assume a circular-equatorial orbit for the third body, in this work it is assumed the Earth follows a Keplerian orbit defined by its inclination $I_\oplus$, longitude of the ascending node $\Omega_\oplus$, and true latitude $u_\oplus = \omega_\oplus + f_\oplus$ measured with respect to the lunar equatorial plane.

Under these conditions, the cosine of the angle $S$ is derived from the dot product $\mathbf{r} \cdot \mathbf{r}_\oplus$. By projecting the third body's position vector onto the satellite's orbital frame, $\cos S$ is expressed without simplification as:

\begin{equation}
	\cos S = \mathcal{A} \cos(\omega + f) + \mathcal{B} \sin(\omega + f)
	\label{eq:cosS_AB}
\end{equation}

\noindent where the coefficients $\mathcal{A}$ and $\mathcal{B}$ encapsulate the geometric coupling between the two orbits:

\begin{equation}
	\begin{aligned}
		\mathcal{A} &= \cos(\Omega - \Omega_\oplus)\cos u_\oplus + \sin(\Omega - \Omega_\oplus)\sin u_\oplus \cos I_\oplus \\
		\mathcal{B} &= \cos I \left[ -\sin(\Omega - \Omega_\oplus)\cos u_\oplus + \cos(\Omega - \Omega_\oplus)\sin u_\oplus \cos I_\oplus \right] + \sin I \sin u_\oplus \sin I_\oplus
	\end{aligned}
	\label{eq:coeffs_AB}
\end{equation}

To facilitate the expansion in terms of the satellite's true anomaly $f$, the auxiliary variables $\alpha$ and $\beta$ are introduced:

\begin{equation}
	\begin{aligned}
		\alpha &= \mathcal{A} \cos \omega + \mathcal{B} \sin \omega \\
		\beta  &= -\mathcal{A} \sin \omega + \mathcal{B} \cos \omega
	\end{aligned}
\end{equation}

\noindent Substituting Eq. (\ref{eq:cosS_AB}) into Eq. (\ref{eq:V3b_quad}) using these auxiliary variables allows the third-body potential to be written explicitly in orbital elements:

\begin{equation}
	V_{\oplus,{P_2}}=-\frac{1}{2}\frac{\mu_\oplus}{r_\oplus}\left(\frac{a}{r_\oplus}\right)^2 \left(\frac{r}{a}\right)^2 \left(3\alpha^2 \cos^2 f+6\alpha \beta \sin f \cos f+3\beta^2 \sin^2 f -1\right)
	\label{eq:V3b_expanded_alpha_beta}
\end{equation}

\noindent Note that in the expression above, the dependence on the Earth's inclination and node is implicitly contained within the definitions of  $\alpha$ and $\beta$.

\subsection{General form of the Hamiltonian}

To characterize the orbital evolution, the equations of motion are formulated using Delaunay canonical variables. The classical Delaunay action momenta ($L, G, H$) and their conjugate angular coordinates ($l, g, h$) are explicitly defined in terms of the Keplerian orbital elements relative to the selenocentric equatorial frame. Furthermore, to account for the additional degree of freedom introduced by the Earth's periodic motion, this system is extended to an eight-dimensional phase space by including the Earth's mean anomaly $M_{\oplus}$ and its conjugate momentum $T$. This extension effectively renders the Hamiltonian autonomous \citep{nie2019}, yielding the following complete set of variables:

\begin{equation*}
	\begin{aligned}
		L &= \sqrt{\mu a} \quad\quad & l &= M \\
		G &= \sqrt{\mu a (1-e^2)} \quad\quad & g &= \omega \\
		H &= G \cos I \quad\quad & h &=  \Omega \\
		M_{\oplus} & = \omega_{\leftmoon} t + M_{\oplus_0}  \quad\quad & T &= \text{conjugate to }M_{\oplus} \\
	\end{aligned}
\end{equation*}

\noindent where the mean anomaly $M$ is a function of eccentricity and true anomaly, i.e., $M=M(e,f)$. Following the canonical convention established by \cite{brouwer1959}, the equations of motion in this extended phase space are expressed as:

\begin{equation}
	\frac{d}{dt}(L, G, H, M_{\oplus}) = \frac{\partial \mathscr{H}}{\partial (l, g, h, T)}, \quad \quad \frac{d}{dt}(l, g, h, T) = -\frac{\partial \mathscr{H}}{\partial (L, G, H, M_{\oplus})}.
	\label{eq.24}
\end{equation}

The Hamiltonian is expressed as:

\begin{equation}
	{\mathscr{H}} =  \frac{\mu^2}{2 L^2} +  \omega_{\leftmoon}T +  \sum_{n=2}^{\infty} R_n + \sum_{n=2}^{\infty}\sum_{m=1}^{n}R_{nm} + R_{\oplus}
	\label{eq:hamiltonian_total}
\end{equation}

\noindent where 

\begin{equation*}
	\begin{aligned}
		\frac{\mu^2}{2 L^2} \qquad & \text{: represents the Keplerian problem (undisturbed problem)} \\
		\omega_{\leftmoon}T \qquad & \text{: represents the phase-space extension term} \\ 
		\sum_{n=2}^{\infty} R_n \qquad & \text{: represents the disturbing force function associated to the zonal harmonics} \\ 
		\sum_{n=2}^{\infty}\sum_{m=1}^{n}R_{nm} \qquad & \text{: represents the disturbing force function associated to the tesseral and sectorial harmonics} \\ 
		R_{\oplus} \qquad & \text{: represents the third body perturbation (Earth's attraction)}  \\
	\end{aligned}
\end{equation*}

Section \ref{section_truncation} details the truncation of the disturbing function associated with the zonal and tesseral harmonics.  It is important to mention that the disturbing force function is the negative of the corresponding potential, \textit{i.e.}, $R = - V$. Following the perturbation approach of \cite{nie2018} or \cite{saedeleer2006}, the Hamiltonian is organized as:

\begin{equation}
	\mathscr{H} = \mathscr{H}_0 + \mathscr{H}_1 + \mathscr{H}_2
	\label{eq.29}
\end{equation}

\noindent where $\mathscr{H}_0 = \mu^2/(2L^2)$ governs the zeroth-order unperturbed motion, $\mathscr{H}_{1}= \omega_{\leftmoon} T$ represents the phase-space extension term contribution, and $\mathscr{H}_2$ contains all the gravitational perturbations (zonal, tesseral/sectorial, and third-body).

The Hamiltonian depends on three angular variables with distinct timescales: the mean anomaly $l$ (short-period, $\sim$ hours), the node-related quantity $h-M_\oplus$ (medium-period, $\sim$ 27 days), and the argument of pericenter $g$ (long-period, $\sim$ years). Short- and medium-period terms are eliminated by successively applying Hori's method \citep{hori1966}. This procedure is equivalent to the one developed by \cite{giacaglia1970}. It is worth noting that explicitly rewriting the Hamiltonian in terms of Delaunay variables is practically disadvantageous. Since the perturbing potentials are natively expanded in classical orbital elements, an explicit conversion to action variables ($L, G, H$) would yield to complicated expressions. Instead, the canonical operations required by Hori's method—such as partial derivatives and Poisson brackets—can be evaluated via the chain rule mapping between the two coordinate sets. So, to preserve mathematical tractability and physical intuition, all expressions in the next section are written in orbital elements.

\subsection{First application of Hori method: elimination of short-period terms}

A canonical transformation $(L,G,H,M_{\oplus}, l,g,h,T) \rightarrow (L^\ast,G^\ast,H^\ast,M_{\oplus}^\ast, l^\ast,g^\ast,h^\ast, T^\ast)$ is introduced via a generating function $S$ such that the new Hamiltonian $\mathscr{H}^*$ becomes independent of the fast angular variable $l^\ast$. The asterisk superscript ($^\ast$) denotes the new averaged variables.   The new Hamiltonian $\mathscr{H}^\ast$ is expressed implicitly as power series of the small parameter as
\begin{equation}
	\mathscr{H}^\ast = \mathscr{H}_0^\ast+ \sum_{k=1} \mathscr{H}_k^\ast
\end{equation}
\noindent as well as the generating function $\varepsilon S$
\begin{equation}
	\varepsilon S = \sum_{k=0} S_k
\end{equation}
The general algorithm of Hori's method \citep{hori1966} determines the new Hamiltonian and the generating function order by order. The fundamental equations up to soncd order in a small parameter are expressed as:
\begin{align}
	\mathscr{H}_0^\ast  &= \mathscr{H}_0 \label{eq:hori_gen_0} \\
	\mathscr{H}_1^\ast  &=   \mathscr{H}_1 + \{ \mathscr{H}_0, S_1 \}  \label{eq:hori_gen_1} \\
	\mathscr{H}_2^\ast &= \left\{ \mathscr{H}_0,S_2\right\} + \left\{\mathscr{H}_1,S_1\right\} + \frac{1}{2}\left\{  \left\{ \mathscr{H}_0,S_1 \right\},S_1\right\} + \mathscr{H}_2    \label{eq:hori_gen_2} 
\end{align}
\noindent  noindent where $ \mathscr{H}_0$,  $ \mathscr{H}_1$ and $ \mathscr{H}_2$ are, respectively, the zero-order and first-order terms of the Hamiltonian, and,  the symbol $\{ \, , \}$ stands for the Poisson bracket. 
Applying this general algorithm to the current dynamical model,  one finds by Eqs. (\ref{eq:hori_gen_0}) and (\ref{eq:hori_gen_1}), repectively: 
\begin{align}
	\mathscr{H}_0^\ast &= \frac{\mu^2}{2{L^\ast}^2},
	\label{eq:hori_0_applied} \\
	\mathscr{H}_1^\ast &= \omega_{\leftmoon} H^\ast, \label{eq.40} \\
\end{align}
yielding 
\begin{align}
	S_1 &= 0, \\
	\mathscr{H}_2^\ast &= \left\{ \mathscr{H}_0,S_2\right\} + \mathscr{H}_2. \label{eq:hori_gen_2_simp} 
\end{align}
The Poisson bracket simplifies to $\{ \mathscr{H}_0^\ast, S_2 \} = -n^\ast \frac{\partial S_1}{\partial l^\ast}$, where $n^\ast = \mu^2 / {L^\ast}^3$ is the mean motion in the averaged canonical variables. Substituting this into Eq. (\ref{eq:hori_gen_2_simp}), the second-order relation becomes:
\begin{equation}
	\mathscr{H}_2^\ast = -n^\ast \frac{\partial S_2}{\partial l^\ast} + \mathscr{H}_2 
\end{equation}

To eliminate the short-period variations, the new perturbed Hamiltonian $\mathscr{H}_2^\ast$ is chosen to be the averaged value of the perturbing potential over the mean anomaly:

\begin{equation}
	\mathscr{H}_2^\ast = \langle \mathscr{H}_2 \rangle = \frac{1}{2\pi}\int_{0}^{2 \pi}\mathscr{H}_2 \, dl
\end{equation}

\noindent leaving the generating function of this first transformation $S_2$ to absorb the remaining periodic fluctuations:

\begin{equation}
	S_2 = \frac{1}{n^\ast}\int\left[\mathscr{H}_2 - \mathscr{H}_2^\ast\right]dl^\ast
\end{equation}

\subsubsection{Averaged Zonal and Tesseral Potentials}

The averaging process requires integrating functions of the elliptical motion over the mean anomaly.  Utilizing the closed-form eccentricity functions derived in \cite{gagg2025}, the averaged potentials are obtained without series expansion in eccentricity.

For the \textbf{zonal harmonics} $\langle R_n \rangle$ , and  \textbf{tesseral and sectorial harmonics} $\left\langle	R_{nm}  \right\rangle$ , the averaged terms are, respectively:

\begin{equation}
	\langle R_n \rangle  = - \frac{\mu}{a}J_n \left(\frac{a_{\leftmoon}}{a}\right)^n\left(1-e^2\right)^{-\left(n-\frac{1}{2}\right)} \sum_{p=1}^{\frac{n}{2}}F_{n,n-2p}(I)G_{n,n-2p}(e)\cos\left((n-2p)\omega\right),\ \text{for}  \ n~\text{even},
	\label{eq.51}
\end{equation}
\begin{equation}
	\langle R_n \rangle  = - \frac{\mu}{a}J_n \left(\frac{a_{\leftmoon}}{a}\right)^n\left(1-e^2\right)^{-\left(n-\frac{1}{2}\right)} \sum_{p=1}^{ \left[\frac{n}{2}\right]}F_{n,n-2p}(I)G_{n,n-2p}(e)\sin\left((n-2p)\omega\right), \ \text{for}  \ n~\text{odd},
	\label{eq.52}
\end{equation}
\begin{multline}
	\left\langle	R_{nm}  \right\rangle =  \frac{\mu}{a}\left(\frac{a_{\leftmoon}}{a} \right)^n \left(1-e^2\right)^{-\left(n-\frac{1}{2}\right)}  \sum_{p = 1}^{n} 
	F_{n,m,n-2p}(I)  G_{n,n-2p}(e) \left[ C_{nm} \cos\left((n-2p)\omega + m(h-M_\oplus)\right) + S_{nm}\sin\left((n-2p)\omega + m(h-M_\oplus)\right)
	\right] \\ \text{if $n-m$ is even},		
	\label{eq.53}
\end{multline}
\noindent and,
\begin{multline}
	\left\langle	R_{nm}  \right\rangle =  \frac{\mu}{a}\left(\frac{a_{\leftmoon}}{a} \right)^n  \left(1-e^2\right)^{-\left(n-\frac{1}{2}\right)} 	\sum_{p = 1}^{n} 
	F_{n,m,n-2p}(I) G_{n,n-2p}(e) 
	\left[ C_{nm} \sin\left((n-2p)\omega + m(h-M_\oplus)\right)
	- S_{nm}\cos\left((n-2p)\omega + m(h-M_\oplus)\right) 
	\right] \\ \text{if $n-m$ is odd}		
	\label{eq.54}	
\end{multline}
\noindent The functions $G_{n,n-2p}(e)$ are presented in the tables of previous work \citep{gagg2025}. These functions are not the same as those introduced by \citet{kaula1960}, since no expansions in eccentricity are considered in the above results. Note also that Kaula's eccentricity functions are based on Cayley’s tables \citep{cayley}. 

\subsubsection{Averaged Third-Body Potential ($M$-averaged)}

As established, the third-body disturbing potential ($V_{\oplus} = -R_{\oplus}$) is truncated at the second-degree Legendre polynomial, with the corresponding term defined as $R_{\oplus,{P_2}}$. Substituting the expanded form of the third-body potential from Eq. (\ref{eq:V3b_expanded_alpha_beta}) and averaging over the satellite's mean anomaly $l$, 	the single-averaged Hamiltonian is obtained. This eliminates the short-period terms while retaining the dependence on the third body's true position (which is handled in the second averaging).
The resulting expression for the averaged disturbing function $\langle R_{\oplus,{P_2}} \rangle$ is given by:
\begin{multline}
	\langle R_{\oplus,{P_2}} \rangle =\frac{1}{2}\frac{\mu_\oplus}{r_\oplus}\left(\frac{a}{r_\oplus}\right)^2 \Bigg[ 
	\frac{3}{2}(1+4e^2) \left( \alpha_1 + \alpha_2 \cos(2u_\oplus) + \alpha_3 \sin(2u_\oplus) \right) 
	+ \frac{3}{2}(1-e^2) \left( \beta_1 + \beta_2 \cos(2u_\oplus) + \beta_3 \sin(2u_\oplus) \right) 
	- \left(1 + \frac{3}{2}e^2\right) \Bigg]
	\label{eq:R3b_single_avg}
\end{multline}
\noindent where $u_\oplus = (\omega_\oplus + f_{\oplus})$ is the true latitude of the Earth. The coefficients $\alpha_i$ and $\beta_i$ capture the coupling between the satellite's argument of pericenter $\omega$ and the geometry of the third body:

\begin{subequations}
	\begin{align}
		\alpha_1 &= \frac{1}{4}(b_1^2 + b_2^2 + c_1^2 + c_2^2) + \frac{1}{4}(b_1^2 + b_2^2 - c_1^2 - c_2^2)\cos(2\omega) + \frac{1}{2}(b_1 c_1 + b_2 c_2)\sin(2\omega) \\
		\alpha_2 &= \frac{1}{4}(b_1^2 - b_2^2 + c_1^2 - c_2^2) + \frac{1}{4}(b_1^2 - b_2^2 - c_1^2 + c_2^2)\cos(2\omega) + \frac{1}{2}(b_1 c_1 - b_2 c_2)\sin(2\omega) \\
		\alpha_3 &= \frac{1}{2}(b_1 b_2 + c_1 c_2) + \frac{1}{2}(b_1 b_2 - c_1 c_2)\cos(2\omega) + \frac{1}{2}(b_1 c_2 + b_2 c_1)\sin(2\omega)
	\end{align}
\end{subequations}
\noindent and,
\begin{subequations}
	\begin{align}
		\beta_1 &= \frac{1}{4}(b_1^2 + b_2^2 + c_1^2 + c_2^2) - \frac{1}{4}(b_1^2 + b_2^2 - c_1^2 - c_2^2)\cos(2\omega) - \frac{1}{2}(b_1 c_1 + b_2 c_2)\sin(2\omega) \\
		\beta_2 &= \frac{1}{4}(b_1^2 - b_2^2 + c_1^2 - c_2^2) - \frac{1}{4}(b_1^2 - b_2^2 - c_1^2 + c_2^2)\cos(2\omega) - \frac{1}{2}(b_1 c_1 - b_2 c_2)\sin(2\omega) \\
		\beta_3 &= \frac{1}{2}(b_1 b_2 + c_1 c_2) - \frac{1}{2}(b_1 b_2 - c_1 c_2)\cos(2\omega) - \frac{1}{2}(b_1 c_2 + b_2 c_1)\sin(2\omega)
	\end{align}
\end{subequations}
\noindent The auxiliary geometric parameters $b_i$ and $c_i$ depend solely on the relative orientation of the orbital planes (inclinations $I, I_\oplus$ and the nodal difference $\Delta\Omega = \Omega - \Omega_\oplus$):
\begin{equation}
	\begin{aligned}
		b_1 &= \cos(\Delta\Omega) \\
		b_2 &= \sin(\Delta\Omega)\cos I_\oplus \\
		c_1 &= -\cos I \sin(\Delta\Omega) \\
		c_2 &= \cos I \cos(\Delta\Omega)\cos I_\oplus + \sin I \sin I_\oplus
	\end{aligned}
	\label{eq:geom_coeffs}
\end{equation}
This single-averaged form explicitly retains the Earth's eccentricity (via the term $r_\oplus$) and the full inclination geometry (via $b_i, c_i$). The presence of $u_\oplus$ indicates that the potential still fluctuates with the period of the Earth's orbit (medium period), requiring a second application of Hori's method to isolate the mean dynamics.

\subsubsection{Computation of the Generating Function $S_2$}

The generating function $S_2$, associate to the short-period terms (dependent on the mean anomaly $l^*$), is determined by the indefinite integral of the periodic part of the first-order Hamiltonian. For the zonal harmonics, the generating function is given by:
\begin{equation*}
	S_{2,J_{n}} = \frac{1}{n^\ast} \int \left[R_{n}\right]_{per} dM^\ast =\frac{1}{n^\ast} \left( \int R_{n} dM^\ast - \langle R_{n} \rangle M^\ast\right)
	%	\label{eqp120a}
\end{equation*}
\noindent where the subscript $_{per}$ indicated the periodic contribution. 
Substituting the potential expansions and performing the integration yields the following expressions. 
For {even zonal harmonics}:
\begin{equation}
	\begin{aligned}
		S_{2,J_{2n}}   =& -\sqrt{\mu a^\ast}            J_{2n}\left(\frac{a_{\leftmoon}}{a^\ast}\right)^{2n}\left(1-e^{\ast^2}\right)^{-\left(2n-\frac{1}{2}\right)}  \Bigg\{F_{2n,0}(I^\ast)\left[G_{2n,0}(e^\ast)\left(f^\ast-M^\ast\right)+\sum_{k=1}^{2n-1}G^k_{2n,0}(e^\ast)\sin \left(kf^\ast\right)\right] \\
		&
		+\sum_{p=0}^{n-1}F_{2n,2n-2p}(I^\ast)\left[G_{2n,2n-2p}(e^\ast)\left(f^\ast-M^\ast\right)\cos \left((2n-2p)\omega^\ast\right) \right. \\
		&
		+\left. \sum_{k=1}^{4n-1-2p}G^k_{2n,2n-2p}(e^\ast) \sin \left(kf^\ast+2(n-p)\omega^\ast\right)+\sum_{k=1,p\neq0}^{2p-1}G^k_{2n,-2n+2p}(e^\ast) \sin \left(kf^\ast-2(n-p)\omega^\ast\right) \right] \Bigg\} 
	\end{aligned}
\end{equation}
For \textbf{odd zonal harmonics}:
\begin{equation}
	\begin{aligned}
		S_{2,J_{2n+1}}   =& -\sqrt{\mu a^\ast}            J_{2n+1}\left(\frac{a_{\leftmoon}}{a^\ast}\right)^{2n+1}\left(1-e^{\ast^2}\right)^{-\left(2n+\frac{1}{2}\right)}  \Bigg\{
		\sum_{p=0}^{n}F_{2n+1,2n+1-2p}(I^\ast)\left[G_{2n+1,2n+1-2p}(e^\ast)\left(f^\ast-M^\ast\right)\sin \left((2n+1-2p)\omega^\ast\right) \right. \\
		&
		-\left. \sum_{k=1}^{4n+1-2p}G^k_{2n+1,2n+1-2p}(e^\ast) \cos \left(kf^\ast+(2n+1-2p)\omega^\ast\right)+\sum_{k=1,p\neq0}^{2p-1}G^k_{2n+1,-2n-1+2p}(e^\ast) \cos \left(kf^\ast-(2n+1-2p)\omega^\ast\right) \right] \Bigg\} 
	\end{aligned}
\end{equation}
\noindent where the eccentricity functions $	G_{2n,0}^k(e^\ast)$, $G_{2n,2(n-m)}^{k}(e^\ast)$ and $G_{2n,2(-n+m)}^{k}(e^\ast)$ are presented in \cite{gagg2025}.
Similarly, the terms of $S_2$ associated with the tesseral and sectorial harmonics ($R_{nm}$) are determined by similar indefinite integral:
\begin{equation*}
	S_{2,J_{nm}} = \frac{1}{n^\ast} \int \left[R_{nm}\right]_{per} dM^\ast =  \frac{1}{n^\ast} \left(\int R_{nm} dM^\ast - \langle R_{nm} \rangle M^\ast\right)
	%	\label{eqp120a}
\end{equation*}
\noindent with $R_{n,m}$ obtained from Eqs \ref{eq.14} or \ref{eq.15} and, $\langle R_{n,m} \rangle$ given by Eqs \ref{eq.53} or \ref{eq.54}. The explicit resulting expressions are omitted here for brevity, as they follow the formulation presented in \cite{gagg2025}; however, care must be taken regarding the redefinition of the Delaunay variables.

Finally, the contribution to $S_2$ associated with the $R_{\oplus,{P_2}}$ is determined by:
\begin{equation*}
	S_{2,R_{\oplus,{P_2}}} = \frac{1}{n^\ast} \int \left[R_{\oplus,{P_2}}\right]_{per} dM^\ast =  \frac{1}{n^\ast} \left(\int R_{\oplus,{P_2}} dM^\ast - \langle R_{\oplus,{P_2}} \rangle M^\ast\right).
	%	\label{eqp120a}
\end{equation*}
Computing the indefinite integral for the third-body interaction results in:
\begin{multline}
	S_{2,R_{\oplus,{P_2}}} = \frac{1}{n^*} \frac{\mu_{\oplus}}{2r_\oplus} \left( \frac{a^*}{r_\oplus} \right)^2 \Bigg\{ \\
	3\alpha^2 \left[ \frac{1}{2}(1+4e^{*2})(E^*-M^*) - \left(\frac{11}{4}e^* + e^{*3}\right)\sin E^* + \frac{1}{2}\left(\frac{1}{2} + e^{*2}\right)\sin(2E^*) - \frac{1}{12}e^*\sin(3E^*) \right] \\
	+ 6\alpha\beta\sqrt{1-e^{*2}} \left[ \frac{5}{4}e^*\cos E^* - \frac{1}{4}(1+e^{*2})\cos(2E^*) + \frac{1}{12}e^*\cos(3E^*) \right] \\
	+ 3\beta^2(1-e^{*2}) \left[ \frac{1}{2}(E^*-M^*) - \frac{1}{4}e^*\sin E^* - \frac{1}{4}\sin(2E^*) + \frac{1}{12}e^*\sin(3E^*) \right] \\
	- \left[ \left(1+\frac{3}{2}e^{*2}\right)(E^*-M^*) - \left(3e^*+\frac{3}{4}e^{*3}\right)\sin E^* + \frac{3}{4}e^{*2}\sin(2E^*) - \frac{1}{12}e^{*3}\sin(3E^*) \right]
	\Bigg\}
	\label{eq:S_3b_generating_function}
\end{multline}

\subsection{Second application of Hori method: elimination of medium-period terms}

After the first canonical transformation, the new Hamiltonian $\mathscr{H}^*$ is expressed as:
\begin{equation}
	\mathscr{H}^*(L^*, G^*, H^*, M_{\oplus}^\ast,\_, g^*, h^*, T^\ast) = \frac{\mu^2}{2{L^\ast}^2} + \omega_{\leftmoon} T^\ast + \mathscr{H}_2^\ast(L^\ast, G^\ast, H^\ast,M_{\oplus},\_, g^\ast, h^\ast,T)
	\label{eq:hamiltonian_star}
\end{equation}
\noindent Since $\mathscr{H}^*$ is independent of the mean anomaly $l^*$, the conjugate momentum $L^*$ (and thus the semi-major axis $a^*$) is a constant of motion. Consequently, the term $\mu^2/(2L^{*2})$ becomes a dynamic constant  and is dropped from the subsequent analysis.

A second canonical transformation $( \_~, G^*, H^*, M_{\oplus}^\ast,\_,  g^*, h^*,T^\ast) \rightarrow ( \_~,G^{**}, H^{**}, \_~,\_~, g^{**}, h^{**},T^{\ast \ast})$ is defined by the generating function $S^\ast$ to eliminate the medium-period terms associated with the Moon's rotation angle $h^* -  M^*_{\oplus}$.  The asterisk superscript ($^{\ast\ast}$) denotes the new double-averaged variables. In this new perturbation step, the term $ \omega_{\leftmoon} T^*$ acts as the unperturbed Hamiltonian of order zero, while the averaged potential $\mathscr{H}_2^*$ acts as the perturbation, thus the homological equation is expressed as:
\begin{equation}
	- \omega_{\leftmoon} \frac{\partial S_1^\ast}{\partial M_{\oplus}^*} + \mathscr{H}_2^{*} = \mathscr{H}_2^{**}
\end{equation}

\noindent which, upon integration with respect to the Earth's mean anomaly, yields the generating function:

\begin{equation}
	S_1^\ast = \frac{1}{\omega_{\leftmoon}} \int (\mathscr{H}_2^* - \mathscr{H}_2^{**}) \, dM_{\oplus}^*
	\label{eq.74}
\end{equation}

\noindent The double-averaged Hamiltonian $\mathscr{H}_2^{**}$ is determined by the averaging principle:
\begin{equation}
	\mathscr{H}_2^{**} = \langle \mathscr{H}_2^* \rangle_{M_{\oplus}^*} = \frac{1}{2\pi} \int_{0}^{2\pi} \mathscr{H}_2^* \, dM_{\oplus}^*
	\label{eq.73}
\end{equation}

\subsubsection{The Mean Hamiltonian $\mathscr{H}_2^{**}$}
The averaging over $M_{\oplus}^*$ completely eliminates the tesseral and sectorial harmonics ($m \neq 0$), as they depend explicitly on the Moon's rotation phase. The zonal harmonics ($m=0$) are already independent of $M_{\oplus}^*$ and remain unchanged, constituting the main mean part of the lunar potential.
Regarding the third-body perturbation, using the single-averaged potential from Eq. \ref{eq:R3b_single_avg} as $\mathscr{H}_2^* $ in Eq. \ref{eq.73} results in the following double-averaged mean term: 
\begin{equation}
	\mathscr{H}_{2_{\oplus,P_2}^{**}} =  \frac{\mu_\oplus}{2a_\oplus} \left( \frac{a^{**}}{a_\oplus} \right)^2 \left(1-e_\oplus^2\right)^{-3/2} \left[ 
	\frac{3}{2}\left(1+4e^{**2}\right)\alpha_1 + 
	\frac{3}{2}\left(1-e^{**2}\right)\beta_1 - 
	\left(1+\frac{3}{2}e^{**2}\right) 
	\right]
\end{equation}

Unlike the simplified model, the inclusion of the Earth's inclination $I_\oplus$ and eccentricity $e_\oplus$ retains a dependence on the satellite's longitude of the ascending node $\Omega^{**}$, explicitly in $\alpha_1$ and $\beta_1$.  As demonstrated in the following sections, this broken axial symmetry is a defining feature of the present theory, as it fundamentally alters the frozen orbit conditions.

The final mean Hamiltonian, combining both lunar and third-body effects, is given by:
\begin{equation}
	\begin{aligned}
		\mathscr{H}_2^{\ast\ast}  = 					
		\mathscr{H}_{2{\oplus,P_2}^{**}} 
		&- \sum_{n=1}^{\infty} \frac{\mu}{a^{\ast\ast} }J_{2n} \left(\frac{a_{\leftmoon}}{a^{\ast\ast} }\right)^{2n}\left(1-{e^{\ast\ast} }^2\right)^{-\left({2n}-\frac{1}{2}\right)} \sum_{p=1}^{n}F_{2n,2n-2p}(I^{\ast\ast} )G_{2n,2n-2p}(e^{\ast\ast} )\cos\left((2n-2p)\omega^{\ast\ast} \right) \\
		& - \sum_{n=1}^{\infty} \frac{\mu}{a^{\ast\ast} }J_{2n+1} \left(\frac{a_{\leftmoon}}{a^{\ast\ast} }\right)^{2n+1}\left(1-{e^{\ast\ast} }^2\right)^{-\left(2n+\frac{1}{2}\right)} \sum_{p=1}^{n}F_{2n+1,2n+1-2p}(I^{\ast\ast} )G_{2n+1,2n+1-2p}(e^{\ast\ast} )\sin\left((2n+1-2p)\omega^{\ast\ast} \right) 
	\end{aligned}
	\label{eq:H3b_double_averaged}
\end{equation}	
The generating function $S_1^\ast$ is calculated by integrating the terms of $\mathscr{H}_2^\ast$ that involves $M_{\oplus}^\ast$ according to Eq. \ref{eq.74}. Below is only highlighted the contribution of the third-body potential since the other terms are the same as determined in \cite{gagg2025}:
\begin{multline}
	S_{1_{\oplus}}^{*} = \frac{1}{\omega_{\leftmoon}} \frac{\mu_\oplus}{2a_\oplus} \left(\frac{a^{**}}{a_\oplus}\right)^2 (1-e_\oplus^2)^{-3/2} \Bigg\{ 
	\frac{3}{2}(1+4e^{**2}) \Bigg[ 
	\left( (f_\oplus-M_\oplus) + e_\oplus\sin f_\oplus \right)\alpha_1 \\
	+ \left( \frac{1}{2}\sin(2f_\oplus+2\omega_\oplus) + \frac{1}{6}e_\oplus(\sin(3f_\oplus+2\omega_\oplus) + 3\sin(f_\oplus+\omega_\oplus)) \right)\alpha_2
	- \left( \frac{1}{2}\cos(2f_\oplus+2\omega_\oplus) + \frac{1}{6}e_\oplus(\cos(3f_\oplus+2\omega_\oplus) + 3\cos(f_\oplus+\omega_\oplus)) \right)\alpha_3 
	\Bigg] \\
	+ \frac{3}{2}(1-e^{**2}) \Bigg[ 
	\left( (f_\oplus-M_\oplus) + e_\oplus\sin f_\oplus \right)\beta_1  \\
	+ \left( \frac{1}{2}\sin(2f_\oplus+2\omega_\oplus) + \frac{1}{6}e_\oplus(\sin(3f_\oplus+2\omega_\oplus) + 3\sin(f_\oplus+\omega_\oplus)) \right)\beta_2
	- \left( \frac{1}{2}\cos(2f_\oplus+2\omega_\oplus) + \frac{1}{6}e_\oplus(\cos(3f_\oplus+2\omega_\oplus) + 3\cos(f_\oplus+\omega_\oplus)) \right)\beta_3 
	\Bigg] \\
	- \left(1 + \frac{3}{2}e^{**2}\right) \left( (f_\oplus-M_\oplus) + e_\oplus\sin f_\oplus \right)
	\Bigg\}
\end{multline}

\subsubsection{Differential equations for the orbital elements}

The time evolution of the mean orbital elements — $a^{**}, e^{**}, I^{**}, \Omega^{**}, \omega^{**}, M^{**}$ — is governed by the system of differential equations derived from the double-averaged Hamiltonian $\mathscr{H}^{**}$. Since the transformation is canonical, these equations are obtained directly from the Poisson brackets involving the orbital elements and the new Haniltonian $\mathscr{H}^{**}$. 
Recall the relationship between the Delaunay variables and the Keplerian elements:

\begin{equation}
	\begin{aligned}
		a^{**} &= \frac{L^{**2}}{\mu}, & M^{**} &= l^{**}, \\
		e^{**} &= \sqrt{1 - \left(\frac{G^{**}}{L^{**}}\right)^2}, & \omega^{**} &= g^{**}, \\
		\cos I^{**} &= \frac{H^{**}}{G^{**}}, & \Omega^{**} &= h^{**}.
	\end{aligned}
	\label{eq:delaunay_relation}
\end{equation}

The Lagrange planetary equations are then derived in terms of the partial derivatives of the Hamiltonian with respect to the orbital elements.
The resulting system is given by:

\begin{align}
	\frac{d a^{**}}{dt} &= 0 
	\label{eq:da_dt} \\
	\frac{d e^{**}}{dt} &= -\frac{\sqrt{1-e^{**2}}}{e^{**}\sqrt{\mu a^{**}}} \frac{\partial\mathscr{H}_2^{**}}{\partial \omega^{**}} 
	\label{eq:de_dt} \\    
	\frac{d I^{**}}{dt} &= \frac{\cot I^{**}}{\sqrt{\mu a^{**}(1-e^{**2})}} \frac{\partial\mathscr{H}_2^{**}}{\partial \omega^{**}} - \frac{1}{\sin I^{**}\sqrt{\mu a^{**}(1-e^{**2})}} \frac{\partial\mathscr{H}_2^{**}}{\partial \Omega^{**}}
	\label{eq:dI_dt} \\
	\frac{d \Omega^{**}}{dt} &= \frac{1}{\sin I^{**} \sqrt{\mu a^{**}(1-e^{**2})}} \frac{\partial \mathscr{H}_2^{**}}{\partial I^{**}}
	\label{eq:dOm_dt} \\
	\frac{d \omega^{**}}{dt} &= \frac{\sqrt{1-e^{**2}}}{e^{**}\sqrt{\mu a^{**}}}\frac{\partial \mathscr{H}_2^{**}}{\partial e^{**}} - \frac{\cot I^{**}}{\sqrt{\mu a^{**}(1-e^{**2})}}\frac{\partial \mathscr{H}_2^{**}}{\partial I^{**}}
	\label{eq:dw_dt} \\
	\frac{d M^{**}}{dt} &= n^{**} - \frac{1-e^{**2}}{e^{**}\sqrt{\mu a^{**}}} \frac{\partial \mathscr{H}_2^{**}}{\partial e^{**}} - 2\sqrt{\frac{a^{**}}{\mu}} \frac{\partial \mathscr{H}_2^{**}}{\partial a^{**}}
	\label{eq:dM_dt}    
\end{align}
\noindent where $n^{**} = \sqrt{\mu/a^{**3}}$ is the mean motion. Note that Eq. (\ref{eq:dI_dt}) explicitly shows the additional term $\frac{\partial\mathscr{H}_2^{**}}{\partial \Omega^{**}}$ which arises from the fact that the third-body geometry breaks the axial symmetry of the system ($I_{\oplus} \neq 0, e_\oplus \neq 0$). If the third body were modeled as circular and equatorial, this term would vanish, recovering the classic result where mean inclination change is zero for a frozen orbit.

%It is important to note that, unlike the ideal zonal problem where the Hamiltonian is independent of the node, the inclusion of the general third-body potential introduces a dependency on $\Omega^{**}$. Consequently, the inclination is not constant for a frozen $\omega^{**}$ but evolves according to the coupled variations of the angular momentum components.

%All partial derivatives of the new Hamiltonian and the generating functions are automatically computed using the Symbolic Math Toolbox in MATLAB for the numerical integration of the system.

\subsection{Solution in orbital elements including the medium- and short-period terms}

The first-order solution for the classical orbital elements including the effects of medium-period terms can be determined by determining the Poisson brackets involving the orbital elements and the generating function $S_1^\ast$, in a similar procedure to the one described in the preceding section. Following Hori's method \citep{hori1966}:

\begin{equation}
	\begin{aligned}
		x^\ast & = x^{\ast\ast} + \left\{x^{\ast\ast},S_1^\ast\right\}.
	\end{aligned}
	\label{eq_app98}
\end{equation}
Similarly, the first-order solution for classical orbital elements including the effects of short-period terms is determined by repeating  the procedure previously described, and using the generating function $S_1$ in the place of $S_2$: 
\begin{equation}
	\begin{aligned}
		x & = x^{\ast} + \left\{x^{\ast},S_2 \right\}		
	\end{aligned}
	\label{eq_app99}
\end{equation}

Despite the similarity with the previous paper from \cite{gagg2025}, special care must be taken regarding the partial derivative $ \frac{\partial {S}_1^\ast}{\partial \Omega^{**}}$, which is no longer zero.

\subsection{Theory validation and Model Truncation \label{section_truncation}}

To define the phase-space validity boundary of the semi-analytical model, the characteristic magnitudes of the Hamiltonian components governing the mean dynamics must be compared. The order of magnitude of the extended phase-space term is given by:
\begin{equation}
	\mathcal{O}\left(\mathscr{H}_1\right) \approx \omega_{\leftmoon}L
	\label{eq:mag_h1}
\end{equation}
\noindent The characteristic magnitudes of the lunar zonal harmonics and the Earth's third-body effect at degrees 2 and 3 are expressed, respectively, as:
\begin{equation}
	\mathcal{O}\left(\mathscr{H}_{J_n}\right) = J_n  \frac{\mu_{\leftmoon}}{a} \left(\frac{a_{\leftmoon}}{a}\right)^n
	\label{eq:mag_Jn}
\end{equation}
\begin{equation}
	\mathcal{O}\left(\mathscr{H}_{\oplus,P_2}\right) = \frac{\mu_{\oplus}}{a_{\oplus}} \left(\frac{a}{a_{\oplus}}\right)^2
	\label{eq:mag_P2}
\end{equation}
\begin{equation}
	\mathcal{O}\left(\mathscr{H}_{\oplus,P_3}\right) = \frac{\mu_{\oplus}}{a_{\oplus}} \left(\frac{a}{a_{\oplus}}\right)^3
	\label{eq:mag_P3}
\end{equation}

\begin{figure}
	\centering
	\includegraphics[width=0.5\linewidth]{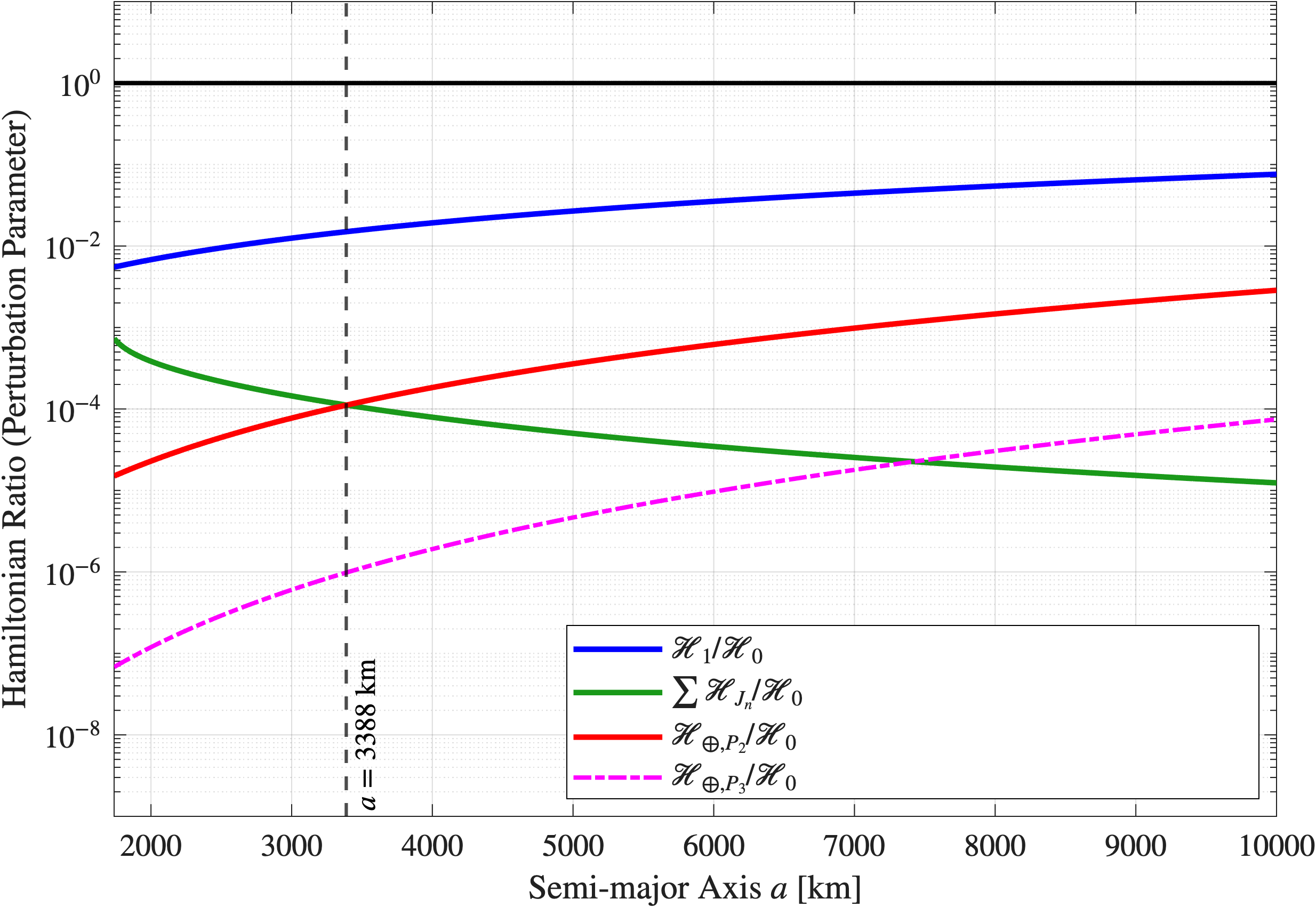}
	\caption{Order of magnitude ration comparision}
	\label{fig:razaohamiltonianasvssemieixo}
\end{figure}

The Lie-Hori perturbation method relies on a hierarchy where the unperturbed Hamiltonian must remain dominant over the sum of the perturbations. 
Figure \ref{fig:razaohamiltonianasvssemieixo} presents the magnitude ratios of these Hamiltonian components relative to $\mathscr{H}_0$ as a function of the semi-major axis, following a structure similar to that presented by \cite{nie2018}.
As expected, the influence of the third-body perturbation increases with the semi-major axis, while the contribution of the lunar zonal harmonics (considered here up to $J_{50}$) decreases. The hierarchy of the model is clearly evidenced by the scaling parameter $\epsilon \approx 10^{-2}$. At the mission's operational altitude, $\mathscr{H}_1/\mathscr{H}_0$ is of the order $\epsilon$, while the second-order third-body term $\mathscr{H}_{\oplus,P_2}/\mathscr{H}_0$ is of order $\epsilon^2$ ($\approx 10^{-4}$), and the third-order term $\mathscr{H}_{\oplus,P_3}/\mathscr{H}_0$ is of order $\epsilon^3$ ($\approx 10^{-6}$).
Notably, for the GARATÉA-L mission ($a = 3388$ km), the magnitudes of the lunar zonal harmonics and the Earth's $P_2$ effect are practically identical, both situated at the $10^{-4}$ level. This fortuitous crossover indicates that, at this specific semi-major axis, the non-spherical gravity field and the primary third-body perturbation contribute equally to the orbital evolution. Furthermore, the crossover between the $P_3$ term and the zonal harmonics occurs at approximately $7500$ km, beyond which the third-body effect becomes the dominant secondary perturbation.

The selection of the truncation degree for both the lunar gravity field and the third-body potential is a critical step in balancing analytical complexity with the fidelity required for mission design. This  next section evaluates the sensitivity of the potential model at the GARATÉA-L operational semi-major axis ($a = 3388$ km).

\subsubsection{Lunar Zonal Harmonics Truncation}

\begin{figure}[h!]
	\centering
	\includegraphics[width=0.5\linewidth]{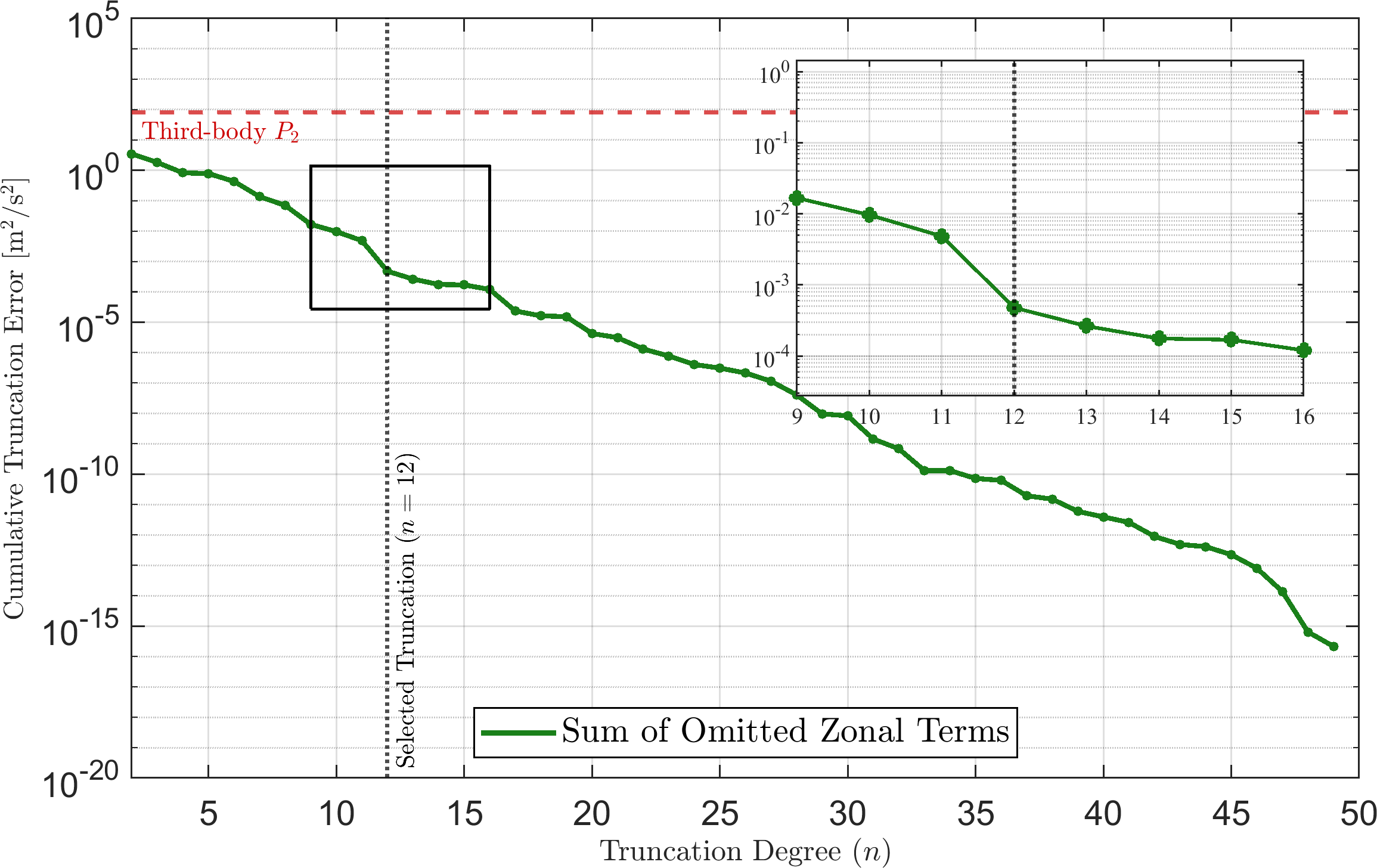}
	\caption{Order of magnitude comparison between Lunar Zonal Harmonics and Earth's Third-Body Perturbation potential terms at $a = 3388$ km.}
	\label{fig:magnitude_analysisZonals}
\end{figure}

Figure \ref{fig:magnitude_analysisZonals} illustrates the cumulative truncation error of the lunar zonal potential, calculated relative to a high-fidelity model including up to $J_{50}$. A sharp ``knee" in the error curve is observed at degree $n=12$. Truncating the model at $J_{11}$ results in a potential error of approximately $3 \times 10^{-3}$ m$^2$/s$^2$. By increasing the degree to $J_{12}$, the error drops to the order of $4 \times 10^{-4}$ m$^2$/s$^2$, representing a one-order-of-magnitude improvement in potential resolution.

Interestingly, the error remains plateaued in the $10^{-4}$ range for truncation degrees between $n=12$ and $n=16$. A further reduction in error (below $10^{-5}$ m$^2$/s$^2$) would only be achieved by extending the expansion beyond $J_{16}$. Therefore, $J_{12}$ represents a good trade-off for first-order semi-analytical theory: it captures the primary geopotential perturbations without introducing the algebraic burden of higher-order terms that offer negligible local gains. In terms of mission impact, an error of $4 \times 10^{-4}$ m$^2$/s$^2$ in the potential ensures that the long-term drift of the mean eccentricity—the primary parameter for minimally variant orbit maintenance—remains within the tolerances of the analytical model hierarchy. For a two years mission, for instance, the error in eccentricity is estimated using Eq. \ref{eq:de_dt}  to be of  $\approx  {1,5 \times 10^{-5}}$ which generated a deviation of only 50 m in pericenter altitude. 

\subsubsection{Third-Body Potential Truncation}

\begin{figure}[h!]
	\centering
	\includegraphics[width=0.5\linewidth]{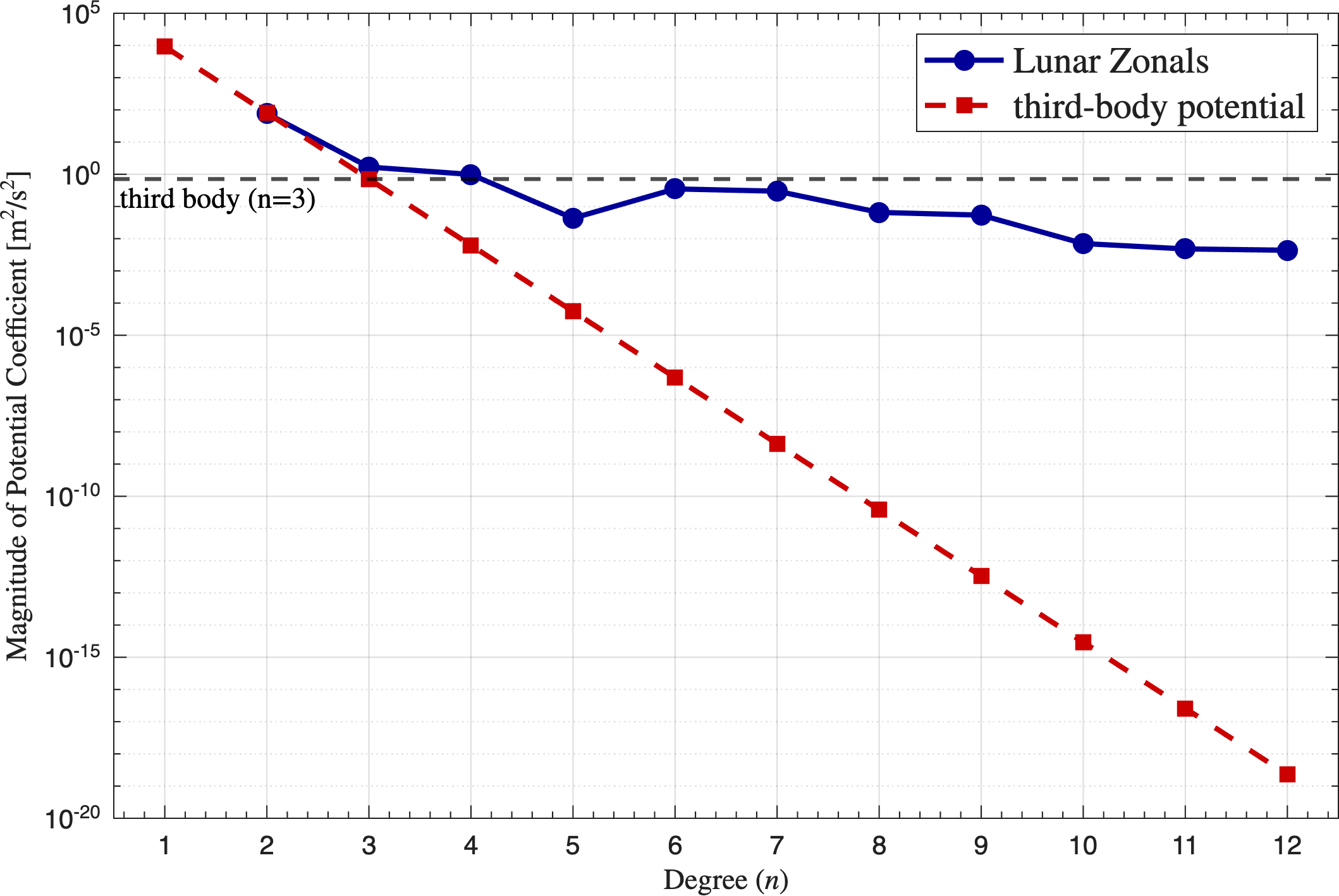}
	\caption{Order of magnitude comparison between Lunar Zonal Harmonics and Earth's Third-Body Perturbation potential terms at $a = 3388$ km.}
	\label{fig:magnitude_analysis}
\end{figure}

The magnitude of the third-body perturbation is governed by the parallax factor $(a/r_{\oplus})$. Figure \ref{fig:magnitude_analysis} compares the individual magnitudes of the lunar zonal coefficients with the degrees of the Earth's third-body potential. Due to the small ratio of the spacecraft's semi-major axis to the Earth-Moon distance ($\approx 3388/384400 \approx 0.0088$), the third-body potential terms decay much faster than the lunar harmonics.

Note that while the $P_2$ term is comparable in magnitude to the primary lunar zonals ($\approx 10^{-2}$ m$^2$/s$^2$), the magnitude of the $P_4$ contribution falls between $J_{11}$ and $J_{12}$, as shown in Fig. \ref{fig:magnitude_analysis}. Because this is on the order of the established truncation limit for the lunar gravity field, the $P_4$ term can be safely neglected without degrading the overall accuracy of the model. This justifies truncating the third-body expansion at $P_2$ or $P_3$ for most applications. However, the inclusion of the third Legendre polynomial ($P_3$) deserves a specific dynamical discussion. To evaluate its secular impact, consider the explicit formulation of the $P_3$ contribution to the Hamiltonian:
\begin{equation}
	\mathscr{H}_{\oplus,P_3}= \frac{\mu_\oplus}{r_\oplus} \left( \frac{r}{r_\oplus} \right)^3 P_3(\cos S)
	\label{eq:R3b_3_def}
\end{equation}
\noindent where $P_3(x) = \frac{1}{2}(5x^3 - 3x)$ and $S$ is the angle between the spacecraft and the Earth. Expanding Eq. (\ref{eq:R3b_3_def}) in terms of the orbital elements yields:
\begin{equation}
	\mathscr{H}_{\oplus,P_3} = \frac{1}{4} \frac{\mu_\oplus}{r_\oplus} \left(\frac{r}{r_\oplus}\right)^3 \Big(\alpha \cos f + \beta \sin f \Big) \left[5 (\alpha^2 - \beta^2) \cos (2f) + 10 \alpha \beta \sin (2f) + 5(\alpha^2 + \beta^2) - 6\right]
	\label{eq:V3b_3_expanded}
\end{equation}
For a circular third-body orbit ($e_\oplus=0$), $\mathscr{H}_{\oplus,P_3}$ vanishes identically upon double averaging due to the odd symmetry of the Legendre polynomial. In the general eccentric-inclined case ($e_\oplus \neq 0, I_\oplus \neq 0$) modeled in this work, this symmetry is broken, allowing a secular contribution to persist. Quantitatively, this residual term is proportional to the product of the parallax factor $(a/a_\oplus)^3$ and the eccentricities $e_\oplus$ and $e$. Given that $e_\oplus \approx 5.5 \times 10^{-2}$, and considering that the parallax factor $(a/a_\oplus)$ introduces a decrease of two orders of magnitude relative to the $P_2$ term, the resulting impact of $P_3$ on the mean dynamics is four orders of magnitude smaller than that of the $P_2$ term. Furthermore, because the fourth-degree Legendre polynomial ($P_4$) is an even function, its double-averaged secular contribution is not penalized by the eccentricities, scaling directly as $\mathcal{O}(\mathscr{H}_{\oplus,P_4}) \sim (a/a_\oplus)^4$. As a result, the eccentricity-penalized $P_3$ term is effectively reduced to the same order of magnitude as the omitted $P_4$ term. While detectable in high-fidelity numerical simulations, this value remains below the intrinsic truncation error of a first-order Lie-Hori theory. Consequently, the third-body dynamics are  well approximated by the eccentric-inclined $P_2$ geometry, which is adopted as the primary model.

\subsection{Frozen orbits}

Stationary solutions for the eccentricity and argument of pericenter, commonly referred to as frozen orbits, are rigorously defined by the conditions $d{e^{**}}/{dt}=0$ and $d{\omega^{**}}/{dt}=0$. However, considering the full third-body perturbation model—accounting for the perturber's eccentricity and inclination — the Hamiltonian $\mathscr{H}_2^{**}$ may retain a dependence on the longitude of the ascending node. Consequently, Eq. \ref{eq:dI_dt} yields a non-zero time derivative for the inclination proportional to $\partial\mathscr{H}_2^{**}/\partial \Omega^{**}$. This secular drift in inclination inevitably induces variations in $\omega^{**}$ and $e^{**}$, preventing the existence of strictly frozen conditions in the long term. As a result, it can be seek `low-drift' orbits: trajectories that vary near the stationary condition of the simplified dynamical model for a specific duration.

%Following the methodology described by \cite{gagg2025}, the conditions for frozen orbits in the simplified model occur when $\omega^{**} = \pi/2$ or $3\pi/2$. It is important to mention that these conditions in only true for the developed perturbation model where the third-body perturbation is put in the same box (order) as the zonal and tesserals. For high altitude orbit, in which the lunar potential harmonics effects becomes irrevalante prevailih the central body, the third-body perturbation becomes larger and the model can be replaced by a three-body problem witch can make surging order families of frozen orbits with other values of $\omega~{**}$.
% To mitigate numerical singularities associated with near-circular orbits ($e^{**} \to 0$), a set of non-singular elements is introduced:

Following the methodology described by \cite{gagg2025}, the conditions for frozen orbits in the circular-equatorial model occur when $\omega^{**} = \pi/2$ or $3\pi/2$. It is important to emphasize that these conditions apply to the dynamical hierarchy of the present perturbation model, wherein the third-body perturbation and the lunar non-spherical harmonics (zonals and tesserals) are assumed to be of the same order of magnitude. For higher-altitude orbits ($a>$ 7500 km), the influence of the lunar gravitational harmonics becomes negligible. In such regimes, the third-body perturbation governs the mean dynamics, consequently, the system should be analyzed within the context of the three-body problem. This shift in the dynamical hierarchy alters the phase-space topology, which can give rise to new families of frozen orbits with stationary conditions at different values of $\omega^{**}$. 

To mitigate numerical singularities associated with near-circular orbits ($e^{**} \to 0$), a set of non-singular elements is introduced:
\begin{equation}
	e_{x}^{**}=e^{**}\cos \omega^{**} \hspace{2cm} e_{y}^{**}=e^{**}\sin \omega^{**}
	\label{eq.100} 
\end{equation} 
The time evolution of these non-singular elements is derived by applying the chain rule to the Lagrange planetary equations, resulting in the following variational equations:
\begin{equation}
	\frac{d{e_{x}^{\ast\ast}}}{dt} = - \frac{\gamma}{\sqrt{\mu a^{\ast\ast}}}\frac{\partial\mathscr{H}_2^{\ast\ast} }{\partial {e_{y}}^{\ast\ast}} + \frac{e_{y}^{\ast\ast}\cot I^{\ast\ast}}{\gamma\sqrt{\mu a^{\ast\ast}}}\frac{\partial \mathscr{H}_2^{\ast\ast} }{\partial I^{\ast\ast}}
	\label{eq.104a}
\end{equation}
\begin{equation}
	\frac{d{e_{y}^{\ast\ast}}}{dt} =  \frac{\gamma}{\sqrt{\mu a^{\ast\ast}}}\frac{\partial\mathscr{H}_2^{\ast\ast} }{\partial e_{x}^{\ast\ast}} - \frac{e_{x}^{\ast\ast}\cot I^{\ast\ast}}{\gamma\sqrt{\mu a^{\ast\ast}}}\frac{\partial \mathscr{H}_2^{\ast\ast} }{\partial I^{\ast\ast}}
	\label{eq.105}
\end{equation}
\noindent with $\gamma=\left(1-{e_{x}^{\ast\ast 2}}-{e_{y}^{\ast\ast 2}}\right)^{\frac{1}{2}}$.
In this non-singular domain, the frozen orbit condition requires the simultaneous vanishing of both rates:
\begin{equation}
	\frac{d{e_{x}^{\ast\ast}}}{dt} = 0 \hspace{2cm}
	\frac{d{e_{y}^{\ast\ast}}}{dt} = 0
	\label{eq.104}
\end{equation}
Due to the symmetries of the averaged Hamiltonian $\mathscr{H}_2^{\ast\ast}$, the partial derivative with respect to $e_{x}^{\ast\ast}$ is proportional to $e_{x}^{\ast\ast}$ itself. Consequently, the right-hand side of Eq. \ref{eq.105} vanishes identically when $e_{x}^{\ast\ast}=0$. This confirms that $e_{y}^{\ast\ast}$ remains constant for configurations where the argument of pericenter is locked at $\pm \pi/2$. Thus, for a fixed semi-major axis $a^{\ast\ast}$ and setting $e_{x}^{\ast\ast}=0$, the problem reduces to finding the roots of the following algebraic equation for $e_{y}^{\ast\ast}$ and $I^{\ast\ast}$:
\begin{equation}
	- \frac{\gamma}{\sqrt{\mu a^{\ast\ast}}}\frac{\partial\mathscr{H}_2^{\ast\ast} }{\partial  e_{y}^{\ast\ast}} + \frac{e_{y}^{\ast\ast}\cot I^{\ast\ast}}{\gamma\sqrt{\mu a^{\ast\ast}}}\frac{\partial \mathscr{H}_2^{\ast\ast} }{\partial I^{\ast\ast}}=0
	\label{eq.107}
\end{equation}
\noindent This non-singular formulation allows for the determination of frozen orbits even in the limit of vanishing eccentricity. Note, however, that the stationarity condition in Eq. \ref{eq.107} is applicable only to the simplified circular-equatorial third-body model ($e_\oplus=I_\oplus=0$), where the satellite's ascending node $\Omega^{\ast\ast}$  does not affect the variational equations of the eccentricity vector (Eqs. \ref{eq.104a} and \ref{eq.104a}).

\subsection{Numerical propagation}

To validate the analytical results and explore the dynamical system beyond the limitations of the averaged models, the Cowell's method \citep{cowell1911} is employed for numerical propagation. This approach consists of the direct numerical integration of the differential equations of motion in Cartesian coordinates, accounting for the summation of all force fields acting on the space vehicle. In the selenocentric inertial frame $Oxyz$ (Fig. \ref{Fig.1}), the equation of motion is given by:
\begin{equation}
	\frac{d^2 \mathbf{r}}{dt^2} = \mathbf{A_0} + \mathbf{A_G} + \mathbf{A_{\oplus}}
	\label{eq.108}
\end{equation}
\noindent where $\mathbf{r}$ is the position of the space vehicle with respect to Moon, $\mathbf{A_0}$ is the absolute acceleration vector of the space vehicle due to gravitational central field of the Moon, $\mathbf{A_G}$ is the absolute acceleration vector of the space vehicle due to higher order and degree terms of the potential field of the Moon, and, $\mathbf{A_{\oplus}}$ is the acceleration due to the third body (Earth)  modeled using the full Newtonian formulation:
\begin{equation}
	\mathbf{A_{\oplus}} = -\mu_\oplus \left[\frac{\mathbf{r}-\mathbf{r_\oplus}}{|\mathbf{r}-\mathbf{r_\oplus}|} + \frac{\mathbf{r_\oplus}}{r_\oplus^3}\right]
\end{equation} 
\noindent where $\mathbf{r_\oplus}$ is the position vector of the Earth relative to the Moon. To evaluate the impact of ephemeris precision on the integration, two distinct strategies for retrieving $\mathbf{r_\oplus}$ are implemented. The first assumes a simplified dynamical model consistent with the semi-analytical theory (eccentric lunar orbit inclined to the equator). The second approach utilizes high-fidelity ephemerides retrieved via the NASA’s Navigation and Ancillary Information Facility (NAIF) SPICE system \citep{spice_cite}. The state vectors of the perturbing bodies are obtained using the Development Ephemeris (DE440) kernel at each integration step.
It is crucial to highlight the fundamental difference between this numerical model and the semi-analytical theory presented earlier. While the semi-analytical approach relies on a second-order Legendre expansion of the disturbing function (truncating higher-order terms), Cowell's method incorporates the exact geometry of the third-body interaction, naturally including all orders of the interaction potential.

\color{black}

%\begin{figure}[htpb]
%	\centering
%	
%	% --- Primeira Subfigura (a) ---
%	\begin{subfigure}[b]{0.4\textwidth}
	%		\centering
	%		% Importação direta: o gráfico adapta-se à largura desta subfigura
	%		\input{Figuras/figura_validade.tex}
	%		\caption{Dynamical hierarchy and validity limit.}
	%		\label{fig:validity_limit}
	%	\end{subfigure}
%	\hfill 
%	% --- Segunda Subfigura (b) ---
%	\begin{subfigure}[b]{0.5\textwidth}
	%		\centering
	%		% Para PNGs, o width=1\textwidth funciona perfeitamente
	%		\includegraphics[width=1\textwidth]{Figuras/figura_validade_contorno.png}	
	%		\caption{Parametric map of the validity boundary.}
	%		\label{fig:validity_contorno}
	%	\end{subfigure}
%	
%	% --- Legenda Global ---
%	\caption{Analysis of the semi-analytical theory validity limit. (a) Perturbation magnitude intersection marking the semi-major axis (56075.81 km) where the third-body effect overtakes the phase-space extension term and $J_2$. (b) Parametric limit across the $(e, I)$ phase-space, showing the exact analytical collapse at polar orbits.}
%	\label{fig:validity_combined}
%\end{figure}

\section{Results}

\begin{figure}[htb!]
	\centering
	\includegraphics[width=0.5\linewidth]{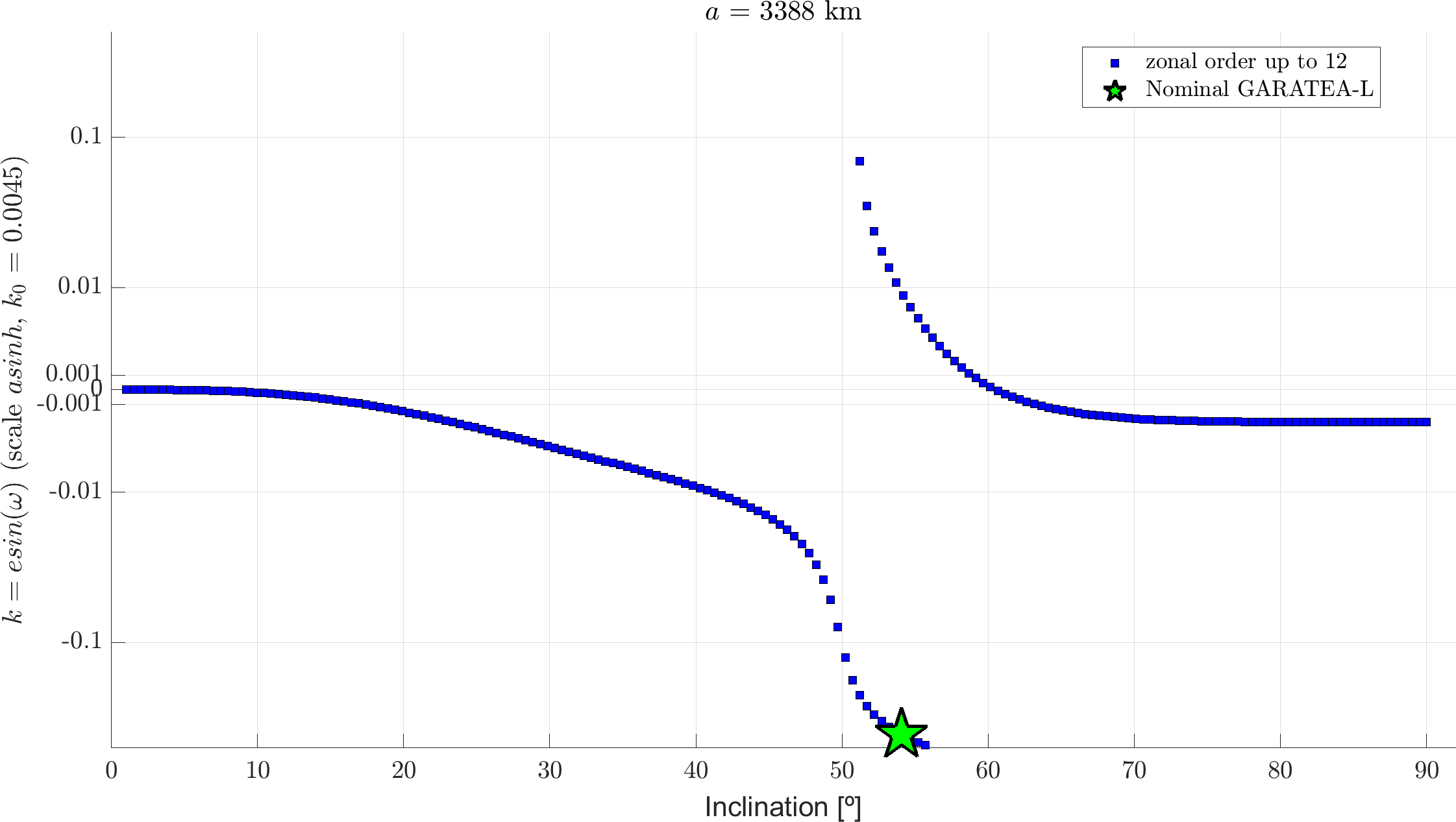}
	\caption{Frozen orbit conditions in models with different zonal harmonics sets and third-body perturbation as a function of inclination, with a = 3388 km.}
	\label{fig_frozencondition}
\end{figure}

The nominal orbital elements for the GARATÉA-L mission were previously established in the COBEM paper \citep{gaggCOBEM2025}, calculated using a semi-analytical model that included zonal harmonics up to degree 12 and an equatorial circular third-body model. For this simplified model, the frozen orbit conditions were identified, and the results are replicated in Fig.  \ref{fig_frozencondition}. This figure is identical to that in the original paper, except for the y-axis, which now uses a symmetric logarithmic (asinh) scale for improved clarity near zero eccentricity. To maintain the pericenter and apocenter altitudes at 300 km and 3000 km, respectively, the specific solution selected by the original authors is marked by a star in the figure. The corresponding mean orbital elements are listed in Tab. \ref{tab:elementos_nominal}.

\begin{center}
	\begin{minipage}{0.4\linewidth}
		%\begin{table}[!h]
		\captionof{table}{Orbital elements for frozen orbit. Zonals to order 12.}
		\centering
		\begin{tabular}{l|c|c}
			\hline
			\textbf{Elements} & \textbf{Mean Elements} \\
			\hline
			$a$ &  3388 km  \\ %2021.598904299798 km & 2.021072649381716e+03
			$e$ &  0.398465 \\ %0.008112932622450 & 0.007803370657573;
			$I$ &  54.045$^\circ$  \\ % 90.483016281033301
			$\omega$ &  270$^\circ$  \\ % 2.655292419171636e+02
			$\Omega$ &  0$^\circ$  \\ % 3.599892113481988e+02
			$M$ &  0$^\circ$\\ % 4.461141907657947
			Pericenter altitude &   300 km \\ %267.1978085995961 %2.673014703727063e+02
			Apocenter altitude &  3000 km \\
			\hline
		\end{tabular}
		\label{tab:elementos_nominal}
		%\end{table}
	\end{minipage}
\end{center}

%\begin{figure}
%	\centering
%	\begin{subfigure} {1.0\linewidth}
	%		\centering
	%		\includegraphics[height=0.18\textwidth]{Figuras/Frozen_SemiEixoExcen_sec_SemiEixo.png}	
	%		\caption{}	
	%	\end{subfigure}
%	\hfill
%	\begin{subfigure} {1.0\linewidth}
	%		\centering
	%		\includegraphics[height=0.18\textwidth]{Figuras/Frozen_SemiEixoExcen_sec_Excen.png}	
	%		\caption{}	
	%	\end{subfigure}
%	\hfill
%	\begin{subfigure} {1.0\linewidth}
	%		\centering
	%		\includegraphics[height=0.18\textwidth]{Figuras/Frozen_SemiEixoExcen_sec_Inclinacao.png}	
	%		\caption{}	
	%	\end{subfigure}
%	\hfill
%	\begin{subfigure} {1.0\linewidth}
	%		\centering
	%		\includegraphics[height=0.18\textwidth]{Figuras/Frozen_SemiEixoExcen_sec_Argumento.png}	
	%		\caption{}	
	%	\end{subfigure}
%	\begin{subfigure} {1.0\linewidth}
	%		\centering
	%		\includegraphics[height=0.18\textwidth]{Figuras/Frozen_SemiEixoExcen_sec_Longitude.png}	
	%		\caption{}	
	%	\end{subfigure}
%	\begin{subfigure} {1.0\linewidth}
	%		\centering
	%		\includegraphics[height=0.18\textwidth]{Figuras/Frozen_SemiEixoExcen_sec_AnomaliaMedia.png}	
	%		\caption{}	
	%	\end{subfigure}
%	
%	\caption{ Evolution of the orbital elements and pericenter altitude of the proposed GARATÉA-L orbit for various models over 2 years. Inclination of 54.045$^\circ$.}
%	\label{orbitalElements_apocenter300}
%\end{figure}

\begin{figure}[htpb]
	\centering
	\includegraphics[width=1.0\textwidth]{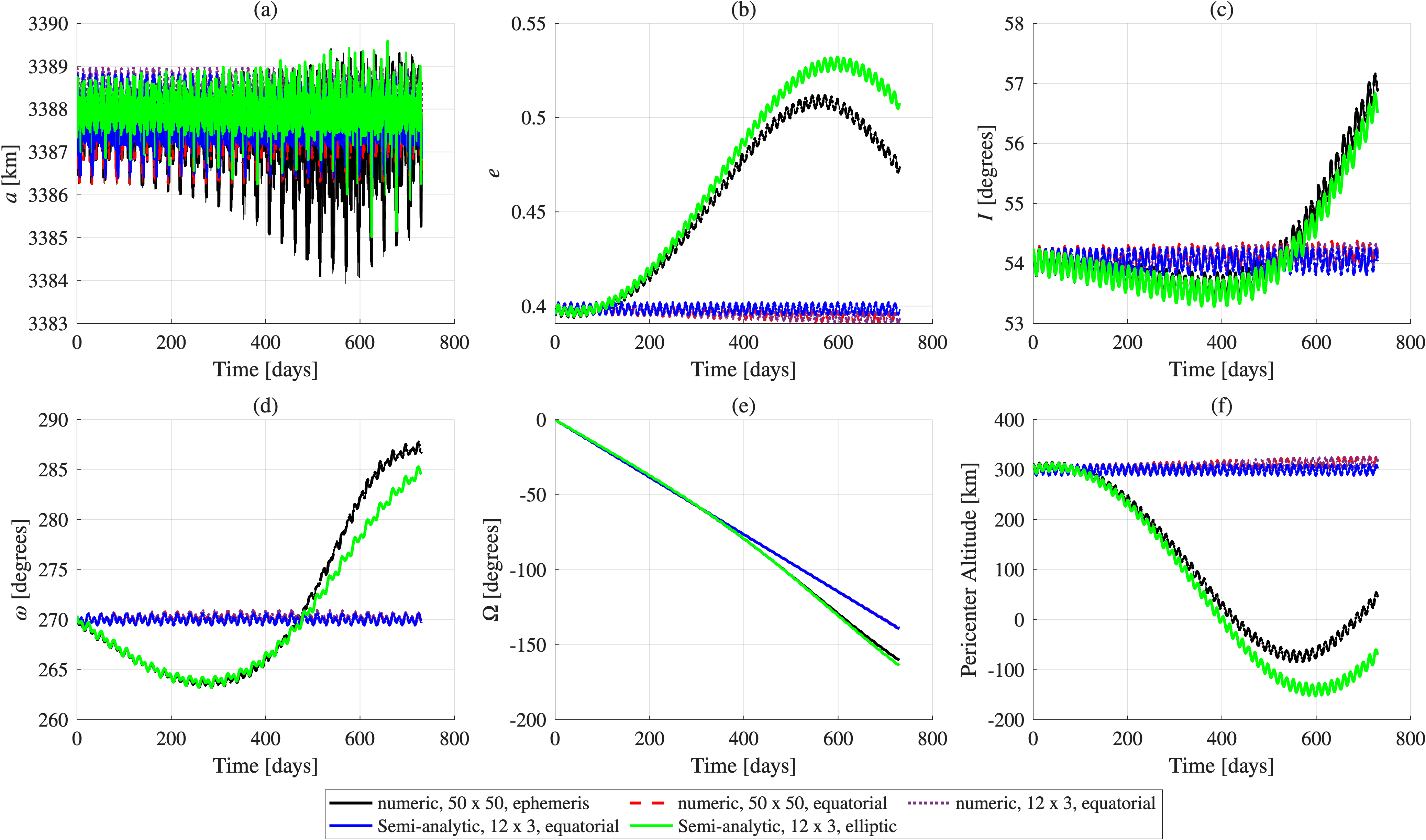}		
	\caption{Evolution of the orbital elements and pericenter altitude of the proposed GARATÉA-L orbit for various models over 2 years. Inclination of 54.045$^\circ$.}
	\label{orbitalElements_apocenter300}
\end{figure}

The authors of the original study then calculated the corresponding osculating elements and performed over a two-year orbital propagation using several models, as shown in Fig. \ref{orbitalElements_apocenter300}. These included a semi-analytical 12$\times$3 model, a numerical propagation 12 $\times$ 3 with circular-equatorial third-body model, a numerical propagation 50 $\times$ 50, and a high-fidelity numerical integration using JPL ephemeris. A key conclusion from that work was that all models showed good agreement except for the one based on ephemeris, which incorporates a more realistic motion of the Earth. This divergence strongly suggested that the simplified circular, equatorial model for the third body is not sufficiently accurate for this mission. Given the orbit's large apocenter and eccentricity, the influence of the Earth's true orbital inclination and eccentricity cannot be neglected. Therefore, the current paper investigates this hypothesis by conducting a new analysis of the GARATÉA-L orbit, considering  the proposed  eccentric-inclined third-body perturbation model. 

Using the same initial osculating elements derived from Tab. \ref{tab:elementos_nominal}, a new propagation is performed with the semi-analytical theory developed in this paper. The result, labeled ``Semi-analytic, 12x3, elliptic", is plotted in Fig. \ref{orbitalElements_apocenter300} for direct comparison. Before discussing these results, it is important to clarify an aspect of the implementation regarding the semi-analytic orbital propagation. Although the third-body perturbation is double-averaged, a specific epoch is required to define the Earth's initial orbital elements. Following the COBEM paper \citep{gaggCOBEM2025}, this epoch is set to  of August 1, 2026. For the simulations presented here, the Earth's subsequent motion is modeled as a simple Keplerian orbit. A more realistic approach, involving retrieving the Earth's position from ephemeris at each time step, is also explored in this work.

The long-term (mean) dynamics are captured through the numerical integration of the variational equations (Eqs. \ref{eq:da_dt}--\ref{eq:dM_dt}) via a 4th-5th order Runge-Kutta algorithm, employing a relative tolerance of the double precision machine ($10^{-16}$). Following the integration, the complete osculating solution is recovered by adding the respective medium-period (Eq. \ref{eq_app98}) and short-period (Eq. \ref{eq_app99}) contributions, as derived from the semi-analytical theory.

\FloatBarrier

An analysis of the new results in Fig. \ref{orbitalElements_apocenter300} shows that including the third body's eccentricity and inclination brings the semi-analytical predictions much closer to the high-fidelity (50 $\times$ 50) ephemeris propagation. The evolution of the semi-major axis (Fig. \ref{orbitalElements_apocenter300}a), eccentricity (Fig. \ref{orbitalElements_apocenter300}b), argument of pericenter (Fig. \ref{orbitalElements_apocenter300}d), and longitude of the ascending node (Fig. \ref{orbitalElements_apocenter300}e) show a strong qualitative match between the new semi-analytical model (green) and the ephemeris model (black). Critically, the pericenter altitude from the semi-analytical model (Fig. \ref{orbitalElements_apocenter300}f) now correctly predicts a collision with the Moon around day 420, a key consequence previously only seen in the ephemeris model. A small discrepancy remains in the inclination (Fig. \ref{orbitalElements_apocenter300}c), which shows a difference of approximately 4 degrees after two years.

Overall, there is good agreement between the semi-analytical 12 $\times$ 3 model with a  more complex third-body perturbation and the numerical 50 $\times$ 50 ephemeris model. However, this agreement reveals a crucial insight: the orbit is no longer frozen. This behavior is clearly illustrated by the evolution of the eccentricity vector: both the argument of pericenter and the eccentricity exhibit significant long-period variations. As seen in the time evolution plots, the argument of pericenter drifts by more than 20 degrees, and the eccentricity grows to a peak near 0.5, which ultimately leads to the predicted collision.

This stark contrast highlights that the previously identified ``frozen" condition is sensitive to the geometric fidelity of the third-body perturbation. In circular-equatorial models, the eccentricity remains bounded near 0.4, preserving a pericenter altitude of 300 km. However, the inclusion of $e_\oplus$ and $I_\oplus$ breaks the symmetry of the problem and triggers a Lidov-Kozai-like exchange between eccentricity and inclination. As eccentricity grows toward 0.52 (Fig.  \ref{orbitalElements_apocenter300}b), the inclination undergoes a secular shift after day 400 (Fig.  \ref{orbitalElements_apocenter300}c), indicating a significant exchange of angular momentum. Furthermore, as the initial inclination resides near the critical region ($54^\circ$–$57^\circ$), the third-body's inclination relative to the Moon's equator breaks the stabilization provided by the $J_n$ terms. 

Consequently, the search for stationary conditions must be revisited. Instead of seeking frozen orbits, the analysis must now focus on identifying ``low-drift" configurations with bounded tolerances in eccentricity and argument of pericenter, utilizing the more complex third-body perturbation model.

\subsection{Low-drift Orbit Condition}

This section revisits the stability analysis for the GARATÉA-L mission ($a = 3388$ km, with pericenter and apocenter altitudes of 300 km and 3000 km, respectively) utilizing the comprehensive $12 \times 3$ semi-analytical theory combined with the eccentric-inclined third-body potential. Finding exact frozen orbits requires solving the stationarity conditions $de_x^{**}/dt = 0$ and $de_y^{**}/dt = 0$ (Eq. \ref{eq.107}). However, unlike in circular-equatorial models, the double-averaged third-body potential now couples the spacecraft's state ($e^{**}$, $I^{**}$, $\Omega^{**}$) with the Earth's orbital geometry ($e_\oplus$, $I_\oplus$, $\Omega_\oplus$). Consequently, solving this system requires prescribing a specific initial value for the satellite's longitude of the ascending node, $\Omega^{**}$. Because $\Omega^{**}$ experiences secular drift, any mathematically stationary state is transient; as the node precesses, the spacecraft can departs from the initial frozen condition. Coupled with the intrinsic time-dependence of the Earth's own orbital elements, a true, static frozen orbit may no longer exist. Therefore, the focus shifts from seeking exact equilibria to identifying ``low-drift" configurations—practical initial conditions that ensure bounded, minimal variations in eccentricity and argument of pericenter over the mission lifetime.

\begin{figure}[htpb]
	\centering
	\includegraphics[width=1.0\textwidth]{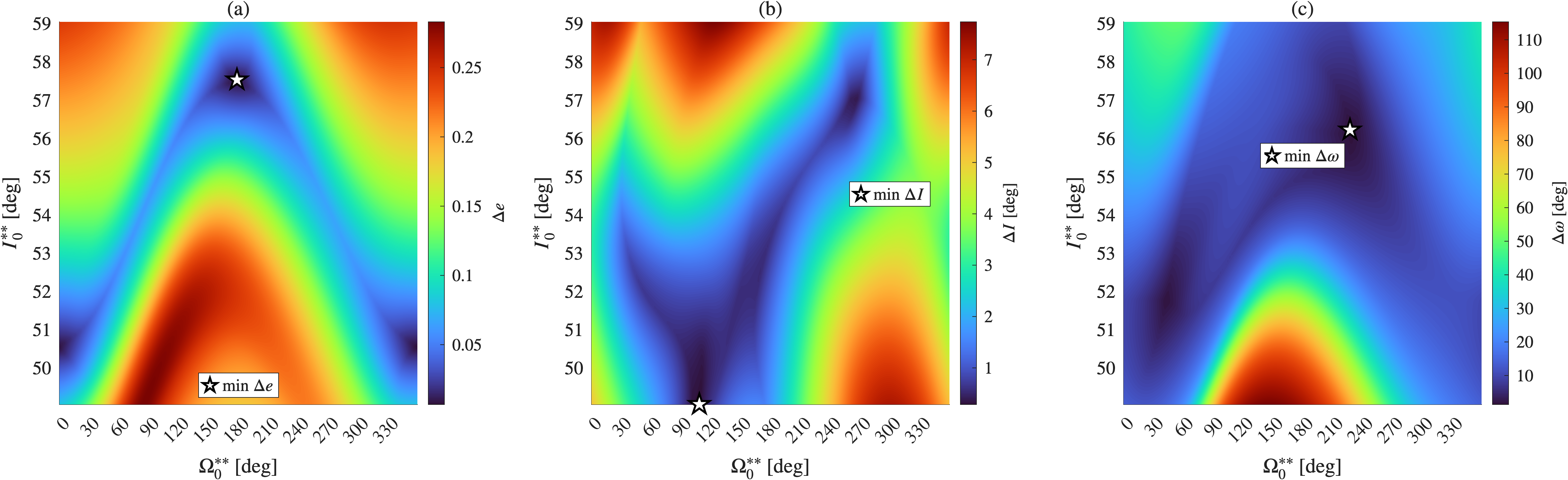}		
	\caption{Parametric study performed with elliptic third-body model around the frozen condition of the circular-equatorial model.}	
	\label{fig_param}
\end{figure}

To find these conditions, a parametric study is performed using the $12 \times 3$ semi-analytical model with the eccentric-inclined third-body potential. A scan around the original frozen condition is conducted, varying the inclination $I^{**}$ from $49^\circ$ to $59^\circ$ (100 points) and the longitude of the node $\Omega^{**}$ from $0^\circ$ to $360^\circ$ (360 points). The resulting 36,000 simulations, which would be computationally expensive via numerical propagation, are completed efficiently thanks to the semi-analytical theory. While a full numerical propagation requires approximately 30 minutes, the analytical approach executes the same task in a few seconds, allowing the entire study to be performed on a standard i3 processor within a practical timeframe. For each simulation, the maximum and minimum values of the orbital elements are recorded to compute their total variation ($\Delta$). Figure \ref{fig_param} consolidates this study through heat maps of $\Delta e$, $\Delta \omega$, and $\Delta I$. The analysis shows that the point of minimum $\Delta I$ lies on the boundary of the studied region, suggesting that a wider scan would be needed if a low-drift condition in inclination is the primary goal. More importantly, the configurations for minimum $\Delta e$ and minimum $\Delta \omega$ do not coincide. From a mission design perspective, priority must be given to the eccentricity, as its growth directly leads to lunar collision. The heat maps reveal that while the region of small $\Delta \omega$ (blue) is predominant, the region of small $\Delta e$ is confined to a narrow, sinusoidal-like stripe. This indicates that the variation of eccentricity is more sensitive to the initial choice of $I^{**}$ and $\Omega^{**}$ than the variation of the pericenter argument. Based on this sensitivity analysis, the set of mean elements that minimizes $\Delta e$ is selected as the new nominal orbit. These values are presented in Table \ref{tab:elementos_new}, where $I^{**}$ and $\Omega^{**}$ are adjusted to $57.53^\circ$ and $178^\circ$, respectively. For this new set, the semi-analytical model predicts an eccentricity variation of approximately $0.0066$, with $\Delta \omega$ and $\Delta I$ staying within $9.665^\circ$ and  $4.239^\circ$, respectively. The following section performs a numerical propagation of this new set to verify if these low-drift conditions hold under higher-fidelity models.

\newpage

\begin{center}
	\begin{minipage}{0.4\linewidth}
		%\begin{table}[!h]
		\captionof{table}{Proposed mean orbital elements. Zonals to order 12.}
		\centering
		\begin{tabular}{l|c|c}
			\hline
			\textbf{Elements} & \textbf{Mean Elements} \\
			\hline
			$a$ &  3388 km  \\ %2021.598904299798 km & 2.021072649381716e+03
			$e$ &  0.398465 \\ %0.008112932622450 & 0.007803370657573;
			$I$ &  57.53$^\circ$  \\ % 90.483016281033301
			$\omega$ &  270$^\circ$  \\ % 2.655292419171636e+02
			$\Omega$ &  178$^\circ$  \\ % 3.599892113481988e+02
			$M$ &  0$^\circ$\\ % 4.461141907657947
			Pericenter altitude &   300 km \\ %267.1978085995961 %2.673014703727063e+02
			Apocenter altitude &  3000 km \\
			\hline
		\end{tabular}
		\label{tab:elementos_new}
		%\end{table}
	\end{minipage}
\end{center}

\FloatBarrier

\subsection{Orbital propagation and analysis}

\begin{figure}[htpb]
	\centering
	\includegraphics[width=1.0\textwidth]{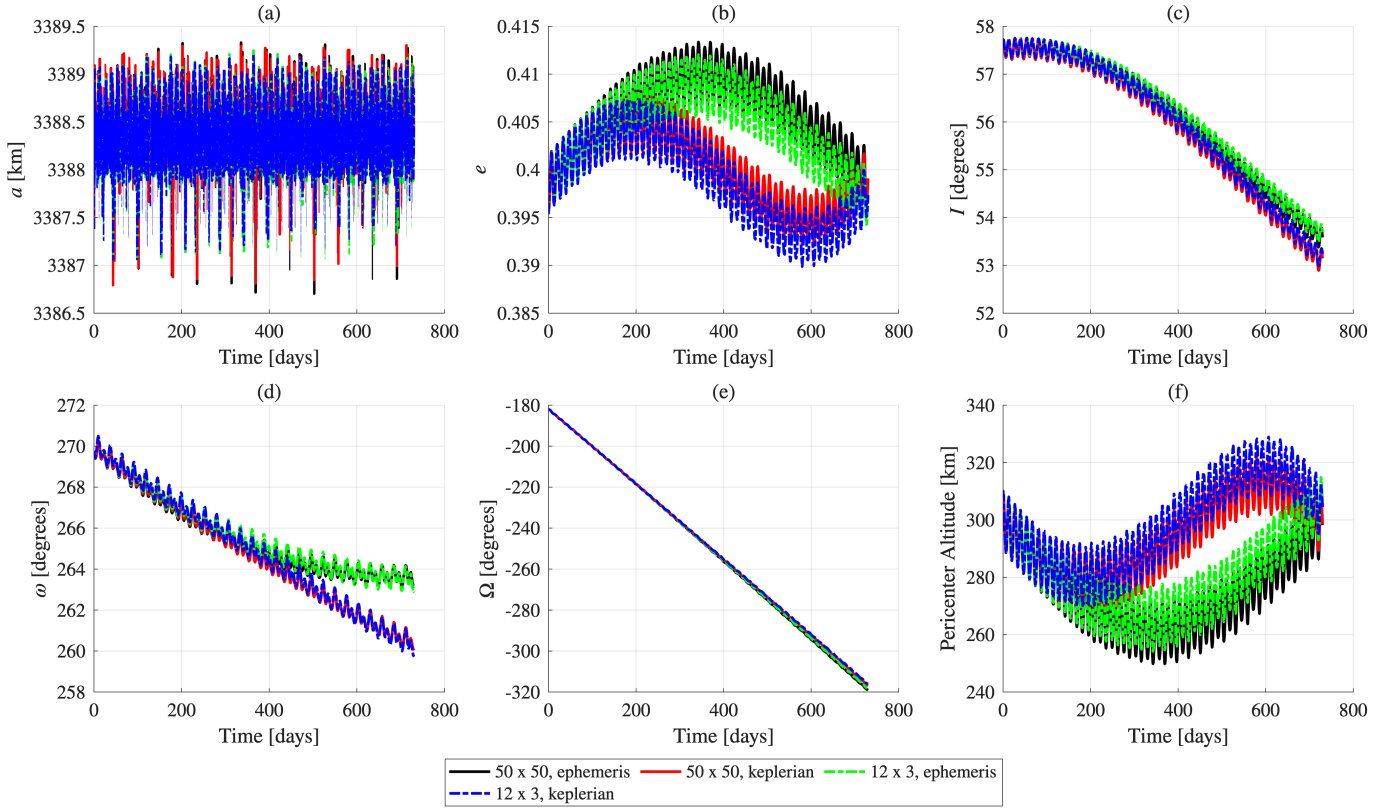}		
	\caption{ Numerical propagation based on Cowell's method.}
	\label{fig_plots_propagacao_2_cowells}
\end{figure}

This section aims to verify whether the newly proposed mean orbital elements successfully generate a low-drift orbit. Furthermore, it seeks to better elucidate the effects of the semi-analytical theory incorporating the non-circular inclined third-body perturbation model and to determine if any significant dynamical effects are still missing from the formulation. Addressing this latter investigation first, Fig. \ref{fig_plots_propagacao_2_cowells} shows the orbital evolution using the numerical propagation to consolidate the model order. The initial orbital elements utilized are the transformed osculating elements corresponding to the proposed mean elements inTable \ref{tab:elementos_new}. Four distinct configurations numerical propagation are considered: regarding the lunar gravitational potential, both a $12 \times 3$ model (including zonal harmonics up to degree $12$ and tesseral harmonics up to degree and order $3$) and a more comprehensive $50 \times 50$ model are analyzed. Regarding the third-body perturbation, the analysis compares a Keplerian motion for the Earth that accounts for its eccentricity $e_\oplus$ and inclination $I_\oplus$ with an ephemeris model in which the position of the third body is retrieved from JPL SPICE ephemeris data at each time step.

It is noted that the most significant differences occur between models with and without ephemeris data. Conversely, the $12 \times 3$ model approximates the $50 \times 50$ configuration very closely, which confirms that the $12 \times 3$ order is an appropriate choice. Regarding the influence of ephemeris, the discrepancies are most visible in the evolution of eccentricity $e$, inclination $I$, argument of pericenter $\omega$, and pericenter altitude (Figs. \ref{fig_plots_propagacao_2_cowells}b, c, d, and f). Nevertheless, these differences remain small over a two-year propagation. For the eccentricity, the maximum variation in the ephemeris models causes $e$ to increase to nearly $0.41$, while in the keplerian model the maximum variation causes $e$ to decreases to nearly $0.39$ from the nominal mean value of $0.398465$, demonstrating that the ``low-drift" behavior is effectively maintained. This higher eccentricity in the ephemeris models culminates in a lower achieved pericenter altitude (nearly $250$ km, Fig. \ref{fig_plots_propagacao_2_cowells}f), which is still close to the nominal $300$ km value and safely distant from a collision scenario.
The argument of pericenter exhibits an expected drift since the established condition prioritizes the minimum $\Delta e$ rather than the minimum $\Delta \omega$.

\def\nome{Frozen\_SemiEixoExcen\_sec\_parametrico}

\FloatBarrier

\begin{figure}[htpb]
	\centering
	\includegraphics[width=1.0\textwidth]{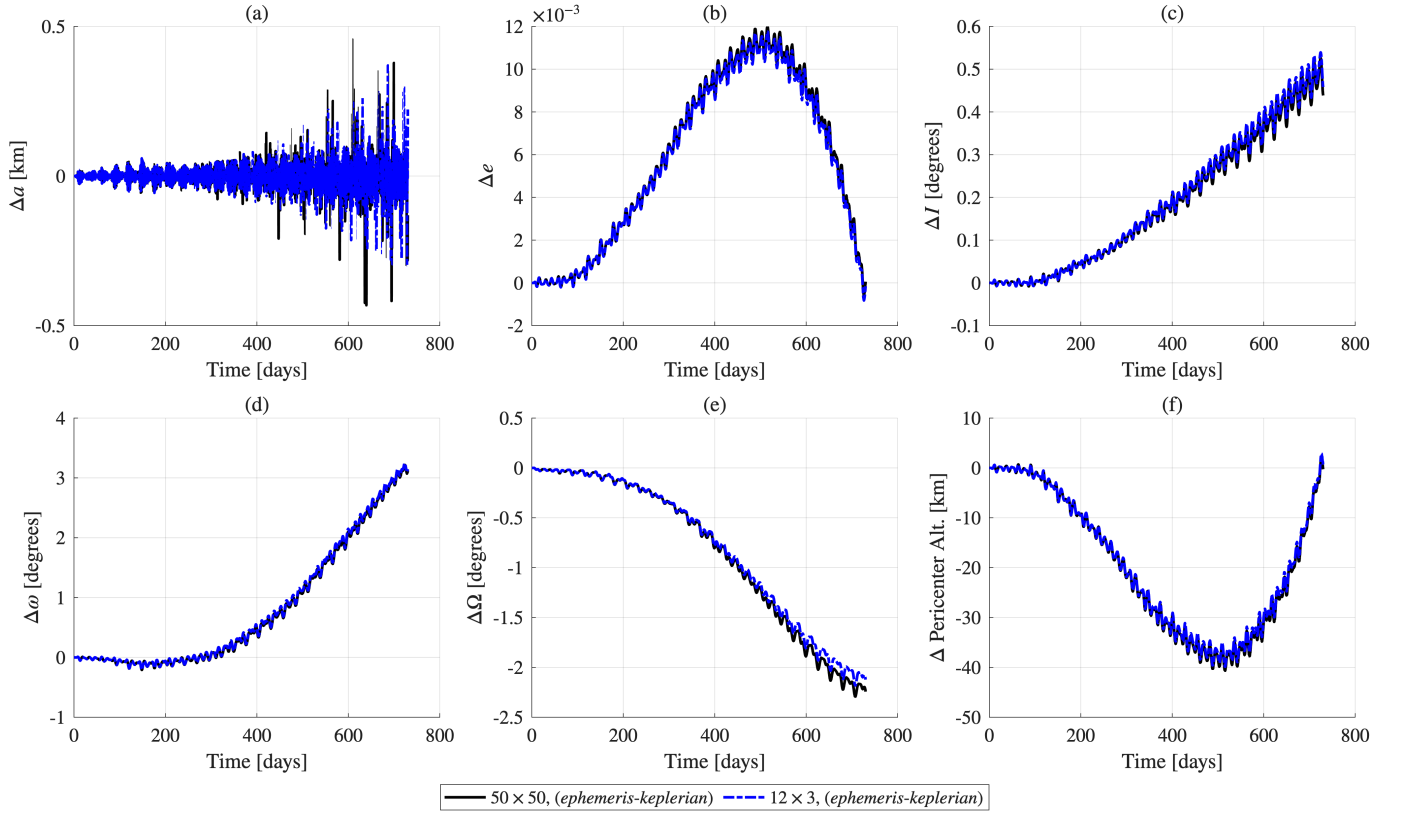}		
	\caption{ Numerical propagation differences.}
	\label{fig_plots_propagacao_2_cowells_diff}
\end{figure}

The discrepancies between the ephemeris-based third-body model and the Keplerian model are further detailed in Fig. \ref{fig_plots_propagacao_2_cowells_diff}, which plots the differences $(\text{ephemeris} - \text{keplerian})$. A primary observation is that the error evolution is virtually identical for both $12 \times 3$ and $50 \times 50$ models, confirming that the dominant source of error resides in the physical modeling of the third body rather than in the degree and order of the lunar gravity field. Interestingly, the errors for some orbital elements do not exhibit a purely cumulative secular growth. For instance, the eccentricity error ($\Delta e$) remains below $0.002$ for the first 100 days, peaks at approximately $0.012$ near day 500, and subsequently returns toward zero by the end of the two-year period. This transient behavior is mirrored in the pericenter altitude difference (Fig. \ref{fig_plots_propagacao_2_cowells_diff}f), which peaks at $40$ km before decreasing. Conversely, the error in the argument of pericenter ($\Delta \omega$) remains negligible for the first year but begins a sustained increase after day 300, reaching approximately $3.2^\circ$ at the end of the mission. Similarly, the longitude of the ascending node ($\Delta \Omega$) and the inclination ($\Delta I$) show a continuous secular divergence, reaching deviations of $2.3^\circ$ and $0.5^\circ$, respectively, after two years. The non-cumulative nature of the errors in $e$ and pericenter altitude suggests a complex geometric coupling, while the steady growth in the orbital plane elements ($\Omega$ and $I$) indicates a persistent shift driven by the non-modeled variations of the Earth's orbit—primarily the solar-induced perturbations on the Earth-Moon system. These results imply that a more precise match with real-world ephemeris would require incorporating the solar influence, which is currently omitted from the semi-analytical theory. Nevertheless, as the overall variations remain within acceptable bounds for preliminary design, the inclusion of the Sun is not addressed in the present study.

\FloatBarrier

\begin{figure}[htpb]
	\centering
	\includegraphics[width=1.0\textwidth]{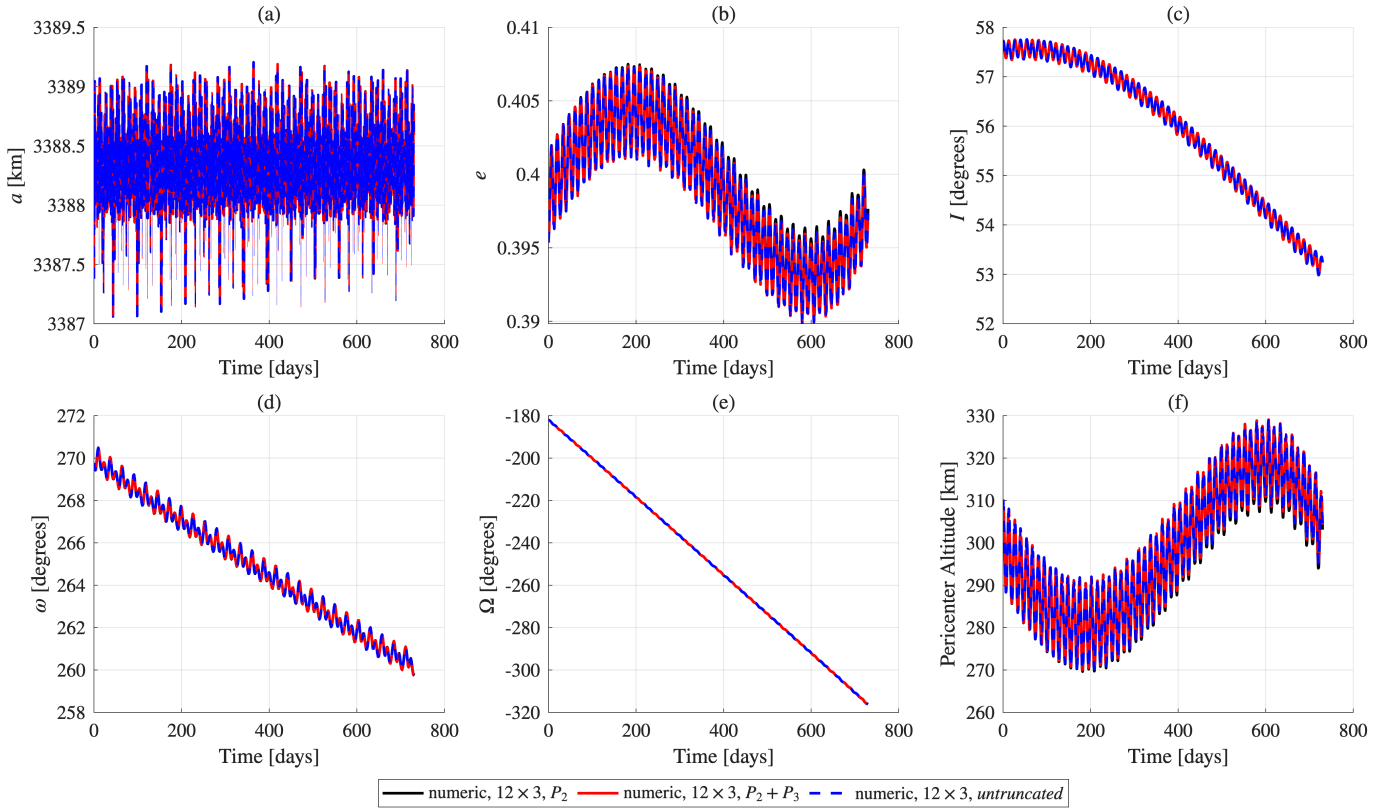}		
	\caption{Long-term evolution of the orbital elements using numerical propagation. The curves compare the effect of truncating the Earth's third-body perturbing function at the second degree ($P_2$), up to the third degree ($P_2+P_3$), and the full un-truncated third-body potential.}
	\label{fig:cowell_elements_truncation}
\end{figure}

\begin{figure}[htpb]
	\centering
	\includegraphics[width=1.0\textwidth]{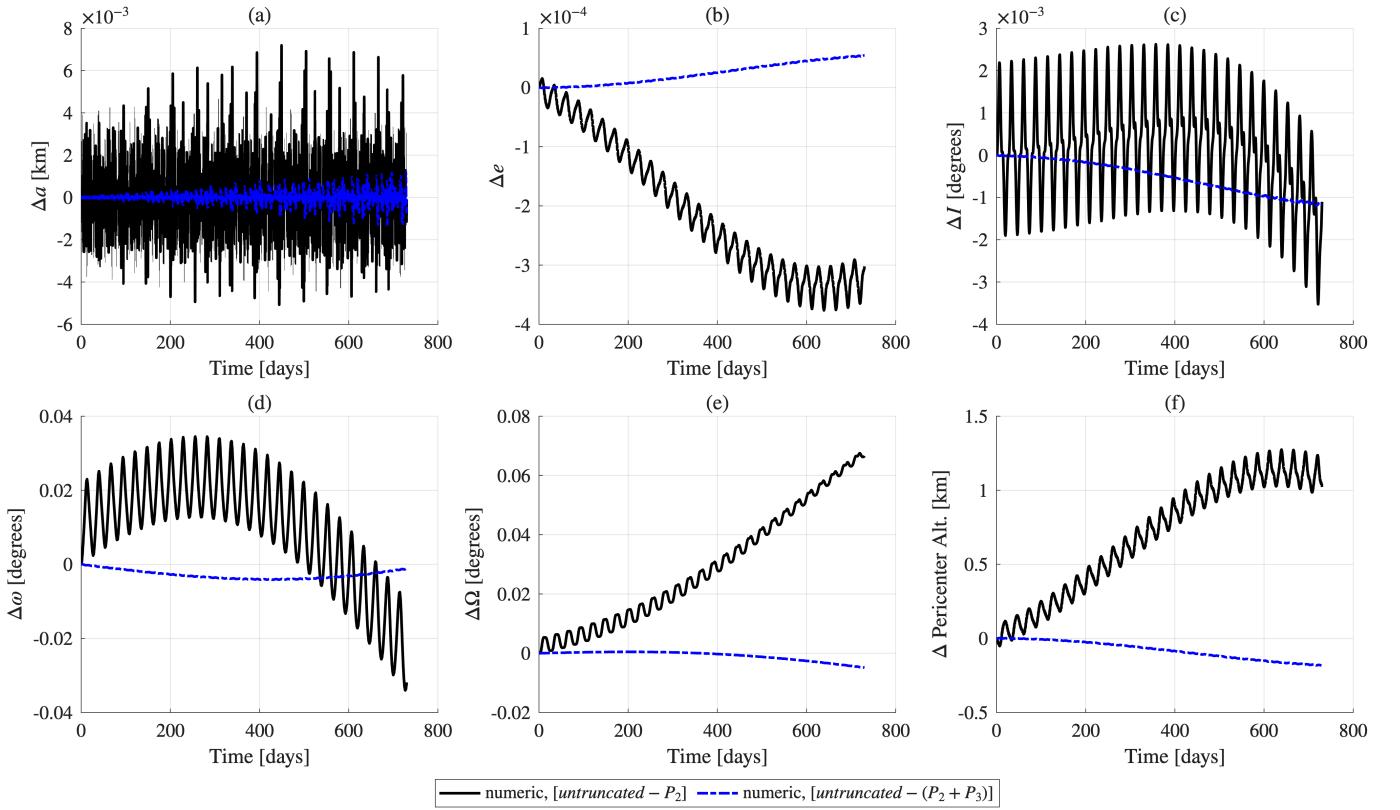}		
	\caption{Differences in the orbital elements caused by the truncation of the Earth's third-body perturbing function. The errors are calculated by comparing the full un-truncated third-body potential against the $P_2$ truncated model and the $P_2+P_3$ truncated model. }
	\label{fig:cowell_diff_truncation}
\end{figure}

Regarding the truncation of the third-body potential, two numerical propagations utilizing Cowell’s method are conducted. These simulations compare the third-body perturbation modeled by a Legendre expansion truncated at the second degree ($P_2$) and the third degree ($P_2 + P_3$) against a benchmark un-truncated model. The orbital evolution for these cases is presented in Fig. \ref{fig:cowell_elements_truncation}, while the corresponding absolute truncation errors are shown in Fig. \ref{fig:cowell_diff_truncation}.

As illustrated in Fig. \ref{fig:cowell_elements_truncation}, the trajectories for the different truncation levels are visually indistinguishable. Although the inclusion of $P_3$ reduces numerical residuals by approximately one order of magnitude, the $P_2$ model maintains high accuracy, with maximum errors of only $1.2$ km in pericenter altitude and $4 \times 10^{-4}$ in eccentricity over a two-year mission. This truncation is physically supported by the dimensionless Hamiltonian ratios in Fig. \ref{fig:magnitude_analysisZonals}, which show that the relative strength of the $P_3$ term ($10^{-6}$) is two orders of magnitude lower than that of the $P_2$ term ($10^{-4}$) at the GARATÉA-L semi-major axis. Therefore, the $P_2$ expansion is considered representative enough for mission design, capturing the dominant secular dynamics while preserving analytical simplicity. Beyond the error magnitude, the characteristic behavior of the curves provides critical insights. The omission of $P_3$ introduces high-frequency oscillations into the error profile, particularly visible in $\Delta I$ and $\Delta \omega$, where the amplitude of the oscillations in the $P_2$ model exceeds the total absolute error of the $P_2 + P_3$ model (see comment  made in the section \ref{section_truncation}.). Furthermore, the error in the longitude of the ascending node ($\Delta \Omega$, Fig. \ref{fig:cowell_diff_truncation}e) exhibits a pronounced secular behavior, with a continuous growth reaching approximately $0.065^\circ$ over two years. This confirms that $P_3$ contributes a measurable secular correction of approximately $0.03^\circ/\text{year}$ to the nodal precession. From a dynamical perspective, this contribution arises from the symmetry breaking induced by the Earth's orbital eccentricity and inclination; unlike circular-equatorial models where odd-degree Legendre polynomials vanish upon double-averaging, the realistic eccentric-inclined geometry prevents the complete cancellation of the third-order potential, introducing a residual torque that modifies the precessional rate of the satellite's orbital plane.

\def\nome{Frozen\_SemiEixoExcen\_sec\_parametrico}

\FloatBarrier

\FloatBarrier

\begin{figure}[htpb]
	\centering
	\includegraphics[width=1.0\textwidth]{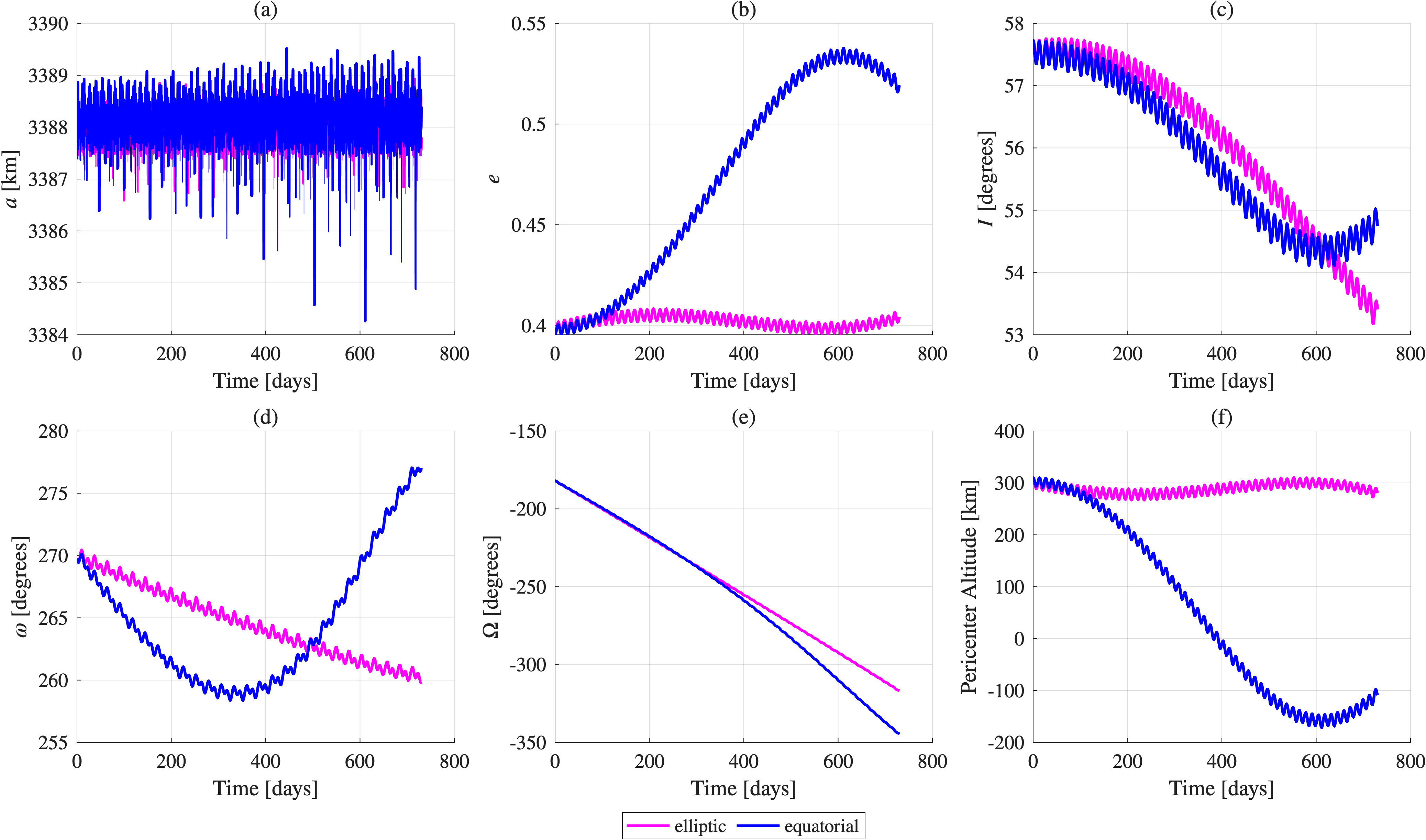}		
	\caption{ Semi-analytic models 12 $\times$ 3: equatorial 3rd-body model and eliptic 3rd-body model.}
	\label{fig_plots_propagacao_3_analiticos}
\end{figure}

Once the $12 \times 3$ lunar gravitational potential model, truncated at the $P_2$ term of the third-body potential, is well establised, the study proceeds to the semi-analytical propagation. Fig. \ref{fig_plots_propagacao_3_analiticos} compares the propagation of the new proposed mean elements utilizing the semi-analytical theory with the ``elliptic" model (including Earth's eccentricity and inclination) against the ``equatorial" model (circular and equatorial third-body model). The results are presented in osculating elements.

\def\nome{plot3}

It should be noted that in the elliptic model, the eccentricity now exhibits a low-drift behavior with small variations around the nominal mean value of $0.398465$ (Fig. \ref{fig_plots_propagacao_3_analiticos}b). Consequently, the pericenter altitude also remains limited, oscillating slightly around $300$ km (Fig. \ref{fig_plots_propagacao_3_analiticos}f). As predicted by the previous analysis, the argument of pericenter in this model presents a minimal drift (Fig. \ref{fig_plots_propagacao_3_analiticos}c), decreasing by only $10^\circ$ over the two-year mission. Conversely, in the equatorial model, these new mean elements no longer correspond to a frozen solution. As a result, the eccentricity (Fig. \ref{fig_plots_propagacao_3_analiticos}b) undergoes large-amplitude long-period variations, reaching peaks of nearly $0.55$. This instability leads to a collision with the Moon around day $390$ of the mission (Fig. \ref{fig_plots_propagacao_3_analiticos}f). The argument of pericenter displays similar long-period instability, dropping to nearly $259^\circ$ and virtually rising to almost $280^\circ$ had the collision not occurred. As expected, a secular drift in the longitude of the ascending node is observed (Fig. \ref{fig_plots_propagacao_3_analiticos}e): up to the point of collision, the models behave similarly; however, over the full two years, the elliptic model shows a total drift of approximately $135^\circ$, closely matching the numerical results presented in Fig. \ref{fig_plots_propagacao_2_cowells}e. Regarding inclination (Fig. \ref{fig_plots_propagacao_3_analiticos}c), the elliptic model yields an evolution historical closer to the numerical results. Finally, both models exhibit only small variations in the semi-major axis around the nominal mean value (Fig. \ref{fig_plots_propagacao_3_analiticos}a).

\FloatBarrier

Despite the inclusion of the eccentricity and inclination of the third body in the semi-analytical model, certain physical effects remain unaccounted for, and specific mathematical approximations are still included. First, regarding the physical model, the secular variations of the Moon's orbital plane and eccentricity — driven primarily by the gravitational influence of the Sun — are not considered. Second, comparing the numerical propagation (with a Keplerian motion of the third body) against the semi-analytical theory, two distinct approximations are present: the truncation of the third-body potential at the second order of the Legendre polynomial ($n=2$), and the intrinsic truncation of Hori's perturbation method, which is developed here as a first-order theory.

To quantify the error introduced by the first-order approximation, an additional propagation is performed using the variational equations. However, instead of propagating the fully averaged mean elements, this analysis integrates the elements $x^{*}$, which result solely from the elimination of short-period terms. By retaining the medium-period terms, the first-order approximation inherent to the second canonical transformation is avoided. Consequently, higher-order coupling effects due to the third body and tesseral harmonics are preserved, albeit at the expense of a higher computational cost.

Figure \ref{fig_plots_erro_1} plots the propagation errors of both the double-averaged semi-analytical theory (two transformations) and the single-averaged theory (one transformation) relative to numerical propagation of equivalent order. The comparison reveals that the single-averaged formulation captures small second-order contributions involving tesseral and third-body couplings that are otherwise eliminated in the double-averaged theory. Consequently, over a two-year mission, the double-averaged model exhibits larger discrepancies in most orbital elements: the eccentricity error is $0.005$ higher than in the single-averaged model, the longitude of the ascending node deviates by nearly $0.7^\circ$, and the pericenter altitude presents an error $20$ km larger. Similarly, the argument of pericenter shows a maximum error increase of approximately $0.4^\circ$. 

%Consequently, over a two-year mission, the double-averaged model exhibits larger discrepancies in most orbital elements: the eccentricity error is $0.005$ higher than in the single-averaged model, the longitude of the ascending node deviates by nearly $0.7^\circ$, and the pericenter altitude presents an error $20$ km larger. Similarly, the argument of pericenter shows a maximum error increase of approximately $0.4^\circ$. 

For the inclination, however, the double-averaged model paradoxically presents a smaller error, decreasing by $0.1^\circ$ compared to the single-averaged results, leaving a residual deviation of approximately $0.2^\circ$ relative to the numerical benchmark. This counter-intuitive behavior is a characteristic of first-order perturbation theories. The single-averaged model retains medium-period terms but still lacks others second-order cross-coupling terms (e.g., the interplay between lunar zonal and third-body effects) required to fully balance them over long integrations. Consequently, the numerical integration of these unbalanced medium-period terms induces a slight artificial secular drift in the inclination. By applying the second averaging procedure, these incomplete medium-period dynamics are entirely filtered out, fortuitously resulting in a more stable secular evaluation and a smaller residual error for the inclination.

\begin{figure}[htpb]
	\centering
	\includegraphics[width=1.0\textwidth]{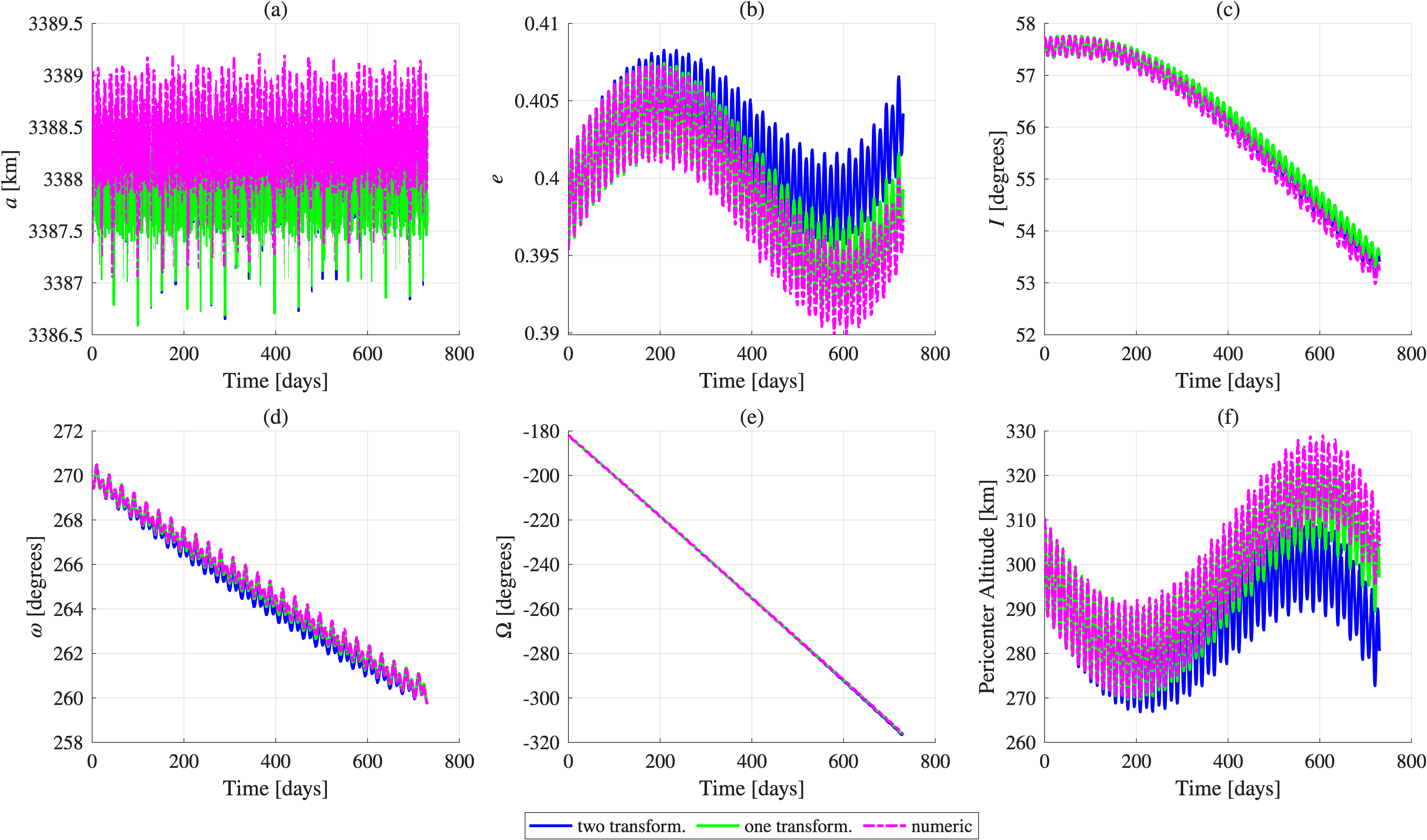}		
	\caption{Orbit's propagation for transformation comparison utilizing the semi-analytic models 12$\times$3 and  Numerical propagation 12$\times$3 with elliptic third-body.}
	\label{fig_plots3b}
\end{figure}

\begin{figure}[htpb]
	\centering
	\includegraphics[width=1.0\textwidth]{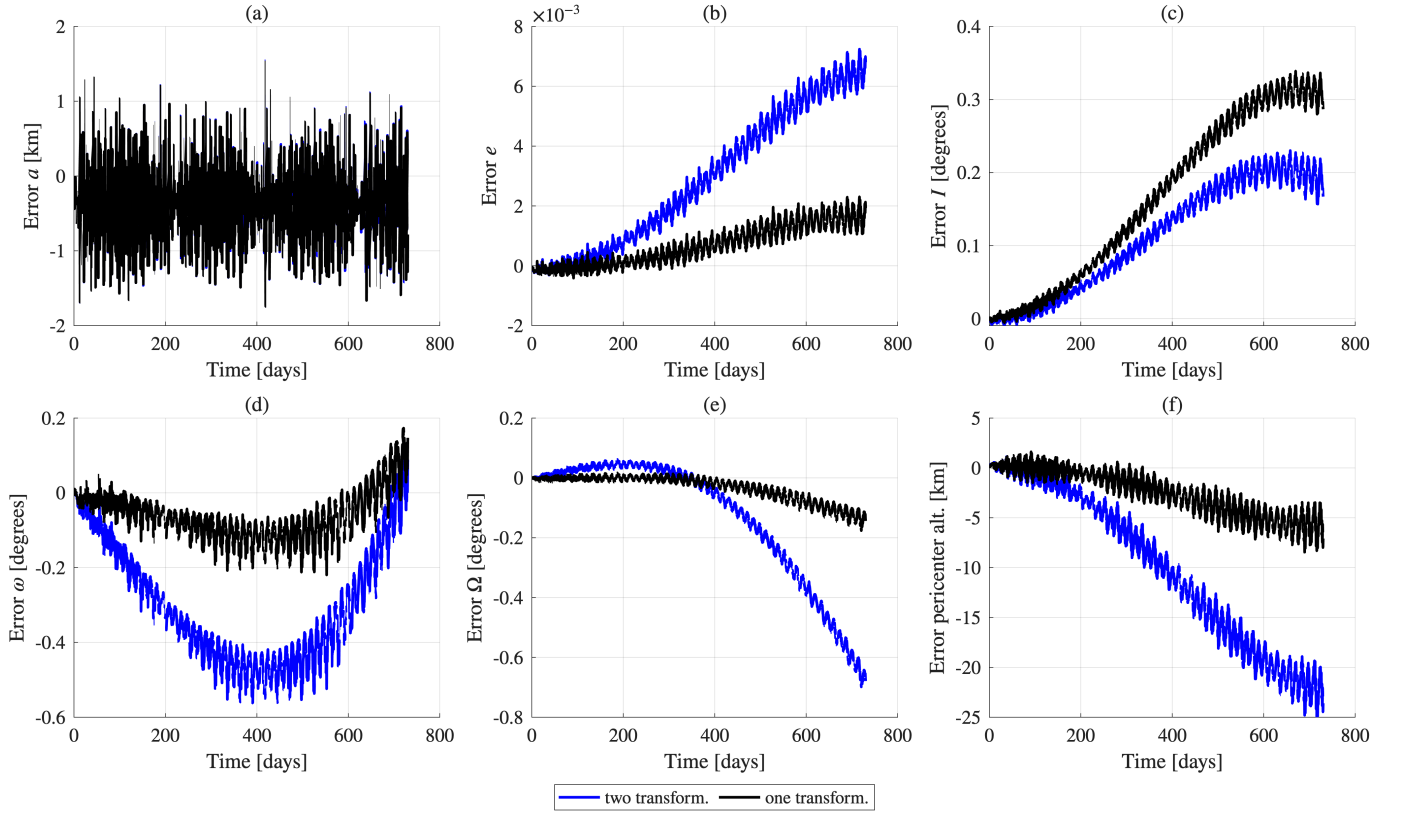}		
	\caption{Transformation comparison. Orbit propagation errors between semi-analytic models 12$\times$3 and Numerical propagation 12$\times$3 with eliptic third-body.}
	\label{fig_plots_erro_1}
\end{figure}

Overall these errors remain small when the propagation of these semi-analytical models is compared to the numerical propagation of the same order, as shown in Fig. \ref{fig_plots3b}. The most significant difference lies principally in the eccentricity error (Fig. \ref{fig_plots_erro_1}b and Fig. \ref{fig_plots3b}b), which necessitates caution regarding the pericenter altitude propagation (Fig. \ref{fig_plots3b}f). It is observable that the higher-order theory (single-averaged) matches the numerical results more closely, leaving only a small residual attributable to the truncation of the third-body perturbation at the second Legendre polynomial ($P_2$). Therefore, the truncation at $P_2$ is deemed sufficiently representative for this analysis. The results suggest that if is required an even more precise theory, the development of a higher-order perturbation theory (or retaining the single-averaged formulation) would be more effective than the inclusion of the third-order Legendre term ($P_3$). However, when comparing these theoretical truncation errors to the discrepancies observed in the ephemeris analysis, a clear hierarchy of importance emerges: improving the physical modeling of the third-body position (to match the ephemeris) is more critical than further developing the analytical order of the theory (observe again the previous comparision shown in Fig. \ref{fig_plots_propagacao_2_cowells}).

\def\nome{plot3b}

\def\nome{plot\_erro1\_refCowell12x3Kepler}

\FloatBarrier

\begin{figure}[htpb]
	\centering
	\includegraphics[width=1.0\textwidth]{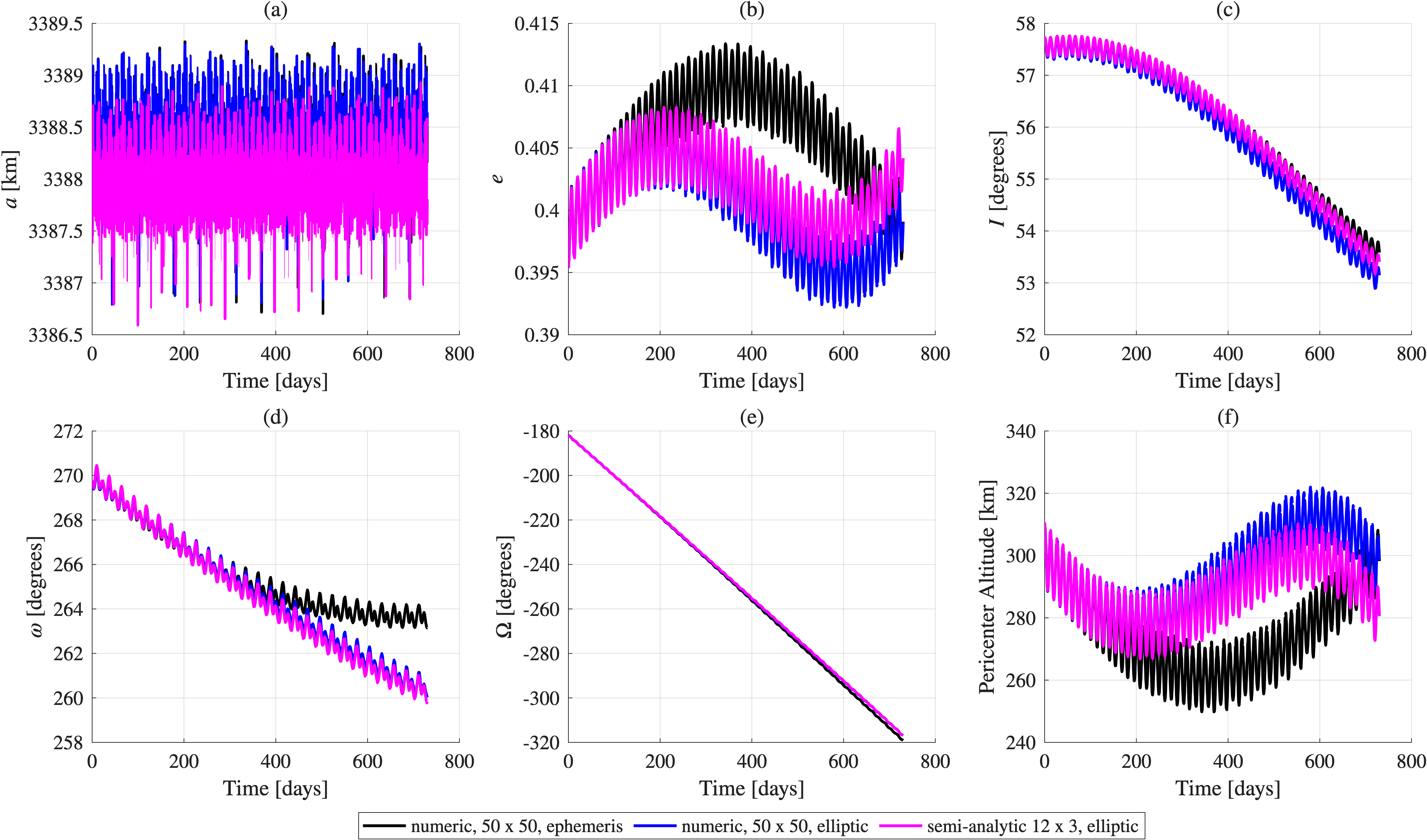}		
	\caption{Orbit propagation. Semi-analytic model 12$\times$3 with eliptic third-body against fidelity models. }
	\label{fig_plots_propagacao_4_conclusao}
\end{figure}

Having consolidated the double-averaged semi-analytical theory within a $12 \times 3$ model—incorporating a third-body potential truncated at the second order while accounting for its eccentricity and inclination—the results are evaluated against high-fidelity numerical benchmarks. The semi-analytical results are compared to a $50 \times 50$ numerical propagation with a Keplerian third-body representation and a $50 \times 50$ numerical model utilizing JPL SPICE ephemeris to retrieve the third-body position at each integration step. The final propagation results are presented in Fig. \ref{fig_plots_propagacao_4_conclusao}.

\begin{figure}[htpb]
	\centering
	\includegraphics[width=1.0\textwidth]{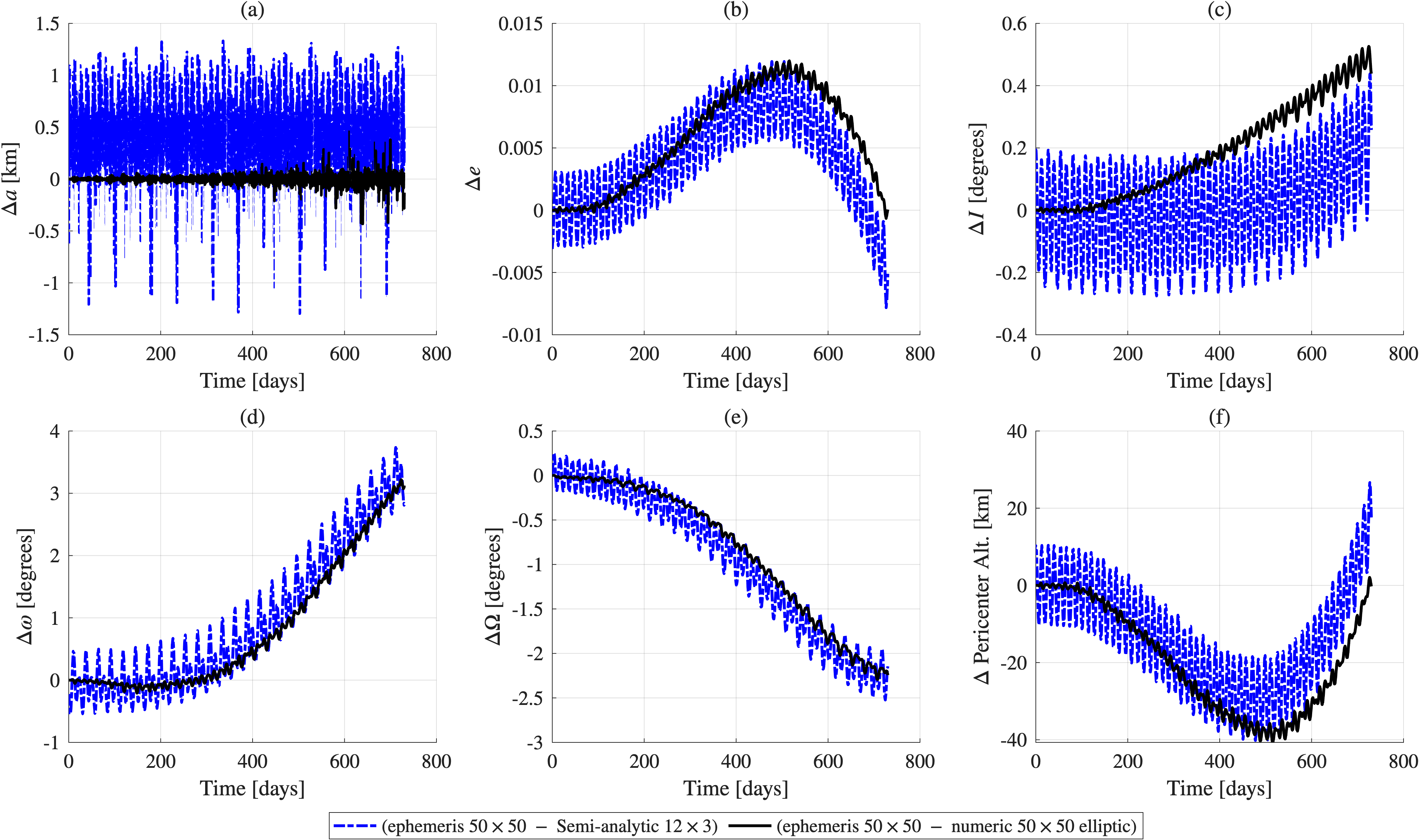}		
	\caption{Orbit propagation. Semi-analytic model 12$\times$3 with eliptic third-body against fidelity models. Differences. }
	\label{fig_plots_propagacao_4_conclusao_diff}
\end{figure}

Overall, the semi-analytical theory demonstrates an excellent match with the Keplerian-based numerical model and strong agreement with the ephemeris-based numerical propagation over the two-year mission duration. The low-drift orbital behavior is clearly evident, as seen in the minimal variations in eccentricity (Fig. \ref{fig_plots_propagacao_4_conclusao}b) and the small secular drift of the argument of pericenter, which shifts from $270^\circ$ to approximately $260^\circ$ in both the semi-analytical and keplerian numerical models (Fig. \ref{fig_plots_propagacao_4_conclusao}d). For this same orbital element, the ephemeris-based results decrease to $265^\circ$.  The eccentricity error ($\Delta e$) and pericenter altitude error exhibit a transient, non-cumulative nature, peaking at approximately $0.012$ and $40$ km, respectively, around day 500 before returning toward zero (Figs. \ref{fig_plots_propagacao_4_conclusao_diff}b and f). In contrast, the angular elements show a steady secular divergence: the argument of pericenter error ($\Delta \omega$) reaches $3.5^\circ$, the longitude of the ascending node ($\Delta \Omega$) deviates by nearly $2.5^\circ$, and the inclination ($\Delta I$) presents a small increase of $0.5^\circ$ over two years. These discrepancies arise from physical effects omitted in the current models, specifically the Sun-induced variations in the Moon's own orbital parameters. Nevertheless, the mission design proves robust, maintaining a safe pericenter altitude throughout the entire duration.

\FloatBarrier

\def\nome{plot4}

\def\nome{plot4}

\FloatBarrier

\section{Conclusion}

This paper details an improved orbit design for the GARATÉA-L Brazilian lunar probe, building upon the initial analysis in previous works. The dynamical model developed here expands the lunar gravitational potential to a $12 \times 3$ harmonics model and, crucially, incorporates a third-body potential that accounts for both the eccentricity and inclination of the Earth's orbit. By applying Hori's method to eliminate short- and medium-period terms, a closed-form semi-analytical solution is obtained, allowing for a deeper understanding of the long-term orbital behavior.

The main motivation for this improvement is the discrepancy found in earlier studies: the nominal frozen orbit—designed using a circular and equatorial model—proved inadequate under realistic conditions. While the simplified model suggested a stationary behavior for the orbital elements, the high-fidelity results show that the spacecraft actually experiences significant eccentricity growth, leading to a lunar collision near day $420$. The semi-analytical theory presented in this work successfully captures this phenomenon, closing the gap between averaged models and numerical ephemeris propagation. Leveraging the theory's computational speed, a parametric study of $36,000$ cases is performed on a standard processor—a task that remains impractical for pure numerical methods. This comprehensive search is necessary to identify ``low-drift" conditions because, in the  more complex model, the mean variational equations explicitly depend on the longitude of the ascending node of the spacecraft. Since this node exhibits a secular drift, a stationary frozen orbit is of difficult to obtain. By adjusting the initial mean elements to $I^{**} = 57.53^\circ$ and $\Omega^{**} = 178^\circ$, a configuration where eccentricity variations are minimized is established. Numerical validation confirms that this new design maintains a pericenter altitude of approximately $250$ km after two years, effectively avoiding the collision scenario seen in previous designs. 

Finally, the analytical and numerical evaluations demonstrate that truncating the third-body potential at the second-degree Legendre polynomial ($P_2$) is mathematically justified. Although the raw magnitude of the third-order term ($P_3$) appears comparable to lower-order lunar zonals, its secular contribution is penalized by the product of the orbital eccentricities ($e e_\oplus$). This penalization reduces its long-term impact to the same order of magnitude as the omitted $P_4$ term, making the $P_2$ truncation a highly robust approximation for this orbital regime. The results indicate that the discrepancies arising from the physical modeling of the Earth's orbit—specifically the solar-induced variations in the Moon's orbital plane—are the dominant source of error, surpassing the effects of analytical truncation. Consequently, future improvements should prioritize accounting for the Sun's influence to better match the ephemeris evolution for this particular value of semi-major axis, rather than focusing on increasing the order of the perturbation theory or expanding the third-body potential. Ultimately, for high-altitude missions like GARATÉA-L, the search for low-drift solutions characterized by minimal orbital drift is a viable path for long-term mission safety.

\section{Acknowledgment}
This study was financed, in part, by the São Paulo Research Foundation (FAPESP), Brasil. Process Number 2024/17018-4.

\bibliographystyle{plainnat} % Substitua o estilo antigo por este\bibliographystyle{jasr-model5-names}
%\biboptions{authoryear}
\bibliography{export}

%--------------------------------------------------------------
\appendix

\end{document}